\documentclass[preprint,12pt]{elsarticle}

\usepackage{amssymb}
\usepackage{amsmath}
\usepackage[table,xcdraw]{xcolor}

\usepackage{xcolor}
\usepackage{subcaption}
\usepackage{adjustbox} 
\usepackage{booktabs}
\usepackage{colortbl}

\usepackage{array} 
\usepackage{wrapfig}
\usepackage{graphicx}
\usepackage{longtable}
\usepackage{lineno}
\usepackage{framed}
\usepackage{float}
\usepackage{rotating}
\usepackage{color}
\usepackage[draft]{todonotes}   
\usepackage{tabularx}
\usepackage{url}
\usepackage{balance}

\usepackage{multirow}
\usepackage{color,soul}

\journal{Forensic Science International: Digital Investigation}

\begin{document}

\begin{frontmatter}




\title{\textit{Breadcrumbs in the Digital Forest}: Tracing Criminals through Torrent Metadata with OSINT}

\author[label1]{Annelies de Jong} 
\ead{anneliesdej2001@gmail.com}

\affiliation[label1]{organization={Tilburg University},
            city={Tilburg},
            country={The Netherlands}}

\author[label1]{Giuseppe Cascavilla} 
\ead{g.cascavilla@tilburguniversity.edu}


\author[label1]{Jessica De Pascale} 
\ead{D.DePascale@tilburguniversity.edu}


\begin{abstract}

This work investigates the potential of torrent metadata as a source for open-source intelligence (OSINT), focusing on user profiling and behavioral analysis. While peer-to-peer (P2P) networks like BitTorrent are well studied in terms of privacy and performance, their metadata is rarely used for investigative purposes. This work presents a proof of concept to show how tracker responses, torrent index data, and enriched IP metadata can reveal patterns linked to high-risk behavior.

The research follows a five-step OSINT process: source identification, data collection, enrichment, behavioral analysis, and presentation of the results. Data were gathered from \textit{The Pirate Bay} and UDP trackers, producing a dataset of more than 60,000 unique IPs across 206 popular torrents. The data were enriched with geolocation, anonymization status, and flags of involvement in child exploitation material (CEM). A case study on sensitive e-books shows how such data can help detect possible interest in illicit content.

Network analysis highlights peer clustering, co-download patterns, and the use of privacy tools by suspicious users. The study shows that torrent metadata, publicly available online, can support scalable and automated OSINT profiling.

This work adds to digital forensics by proposing a new method to extract useful signals from noisy data, with applications in law enforcement, cybersecurity, and threat analysis.

\end{abstract}




\end{frontmatter}



\section{Introduction}\label{chapter:introduction}

In a digital landscape shaped by global connectivity and public data availability, open source intelligence (OSINT) has become an essential tool for collecting and analyzing publicly accessible information. From cybersecurity to law enforcement and digital forensics, OSINT supports investigations by tracing digital behaviors and patterns. However, specific data sources remain underused, including metadata generated by peer-to-peer (P2P) file-sharing activity.


BitTorrent, one of the most widely used P2P protocols, enables decentralized file distribution via small metadata files called torrents. These files contain information such as tracker URLs, file names, and cryptographic identifiers, but do not store content themselves. Although BitTorrent has been extensively studied for piracy detection, network optimization, and protocol analysis \cite{Lareida2018,Gartner2023}, its metadata remains underexplored for user profiling and behavioral analysis in OSINT contexts \cite{PastorGalindo2020,Yadav2023,Torres2024}. Existing research on torrent traffic and IP intelligence has focused on threat detection and infrastructure performance, rather than user profiling or behavioral analysis.


This research addresses that gap by evaluating whether torrent metadata can support scalable and structured OSINT-based user profiling. The study presents a multistage proof of concept that combines torrent index data, tracker responses, IP metadata, and external flagging sources to identify behavioral signals indicative of high-risk activity.


In the context of this research, \textit{high-risk behavior} refers to torrent activity that may suggest increased potential for criminal or harmful intent. Examples include repeated involvement in swarms associated with flagged content (e.g., child exploitation material - CEM), focused interest in sensitive topics (e.g., explosives manuals), and the use of anonymization technologies in combination with such activity. Although these signals do not prove intent or illegality, they can support prioritization of further investigation within OSINT workflows.


In the context of this research, we propose a structured, multi-stage proof of concept that combines torrent index data, tracker responses, IP metadata, and external flagging sources. Although not yet fully automated, the design supports future scalability and aims to address common OSINT challenges like noisy data and the need for cross-source correlation.


The central research question (RQ) that guides this study is the following:

\begin{quote}
\textbf{RQ}: 
To what extent can our proof of concept demonstrate the potential of torrent metadata for OSINT-based user profiling?
\end{quote}


\noindent The main RQ is further explored through three sub-questions (sRQ):

\begin{quote}
\noindent\textbf{sRQ1}: 
What torrent and IP metadata are publicly available and usable for detecting high-risk behavior?
\end{quote}

This question examines what types of data can be accessed from open sources like torrent indexes and trackers. It considers their structure, availability, and potential for identifying suspicious activity.

\begin{quote}
\textbf{sRQ2}: How effectively can our proof of concept map torrent user networks using publicly available data?
\end{quote}


The sRQ2 addresses whether patterns such as co-download behavior or content clustering can be detected through network analysis. The sRQ2 explores how these patterns contribute to profiling and behaviotal insights.

\begin{quote}
\textbf{sRQ3}: 
How suitable are torrent and IP metadata for integration into OSINT workflows in terms of data quality, structure, and processing effort?
\end{quote}

Finally, sRQ3 evaluates the operational viability of using this type of data in real-world OSINT workflows. This includes examining data quality, anonymization challenges, and technical overhead. It offers a broader judgment on whether and how such data can realistically support investigations.


To answer these questions, the study applies a five-stage OSINT workflow: source identification, data collection, processing, analysis, and reporting. sRQ1 relates to finding and collecting useful torrent and IP data. sRQ2 focuses on analysing this data to spot patterns and user behavior. sRQ3 looks at how practical and reliable this data is for real investigations. Together, the questions guide a step-by-step evaluation from data access to real-world use.

Data was collected from The Pirate Bay and public UDP trackers, resulting in a dataset of over 60,000 unique IP addresses across 206 popular torrents. This dataset was enriched with geolocation information, anonymization flags, and external indicators (e.g., associations with child exploitation material).

A focused case study on sensitive e-books demonstrates how metadata can help reveal user interest in high-risk content categories. Network analysis shows patterns of co-downloads, use of privacy-preserving tools, and peer clustering—all of which contribute to behavioral profiling.

Through this approach, the research makes several contributions:

\begin{itemize}
\item 
It introduces \textbf{torrent metadata as a novel and underused OSINT data source} for user profiling. To the best of our knowledge, this is the first structured application of torrent data for this purpose.

\item 
It presents a \textbf{replicable profiling framework} that integrates torrent records, tracker responses, IP intelligence, and third-party flagging into a single workflow.

\item 
It enhances \textbf{signal extraction in noisy environments}, addressing the common challenge of filtering relevant intelligence from large, unstructured data streams.

\item 
It provides tools for \textbf{prioritizing investigations} by identifying behavioral signals that suggest elevated risk or intent.

\item 
Finally, while the proof of concept is manually executed, it lays the foundation for \textbf{future automation and scalability}, enabling potential deployment in real-time OSINT or forensic systems.
\end{itemize}


In conclusion, this research highlights the untapped potential of torrent metadata for behavioral profiling within OSINT workflows. By integrating unconventional data sources into digital investigations, it expands the analytical capabilities available to threat intelligence and law enforcement communities.

\section{Background}\label{chapter:background}


Open source intelligence (OSINT) is the collection and analysis of publicly available data to produce actionable knowledge \cite{Puyvelde2025}. The modern digital world generates vast amounts of information \cite{PastorGalindo2020}, much of which is accessible to anyone, anywhere, at any time. OSINT uses this accessibility to gather, process, and connect data, creating a broader picture of the target. Its strength lies in combining multiple sources to reveal insights that a single stream cannot provide.


OSINT has changed with technology. Early efforts focused on newspapers, public records, and government publications \cite{Puyvelde2025}. The growth of digital platforms, such as social media and online forums, expanded its scope. More recently, big data, automation, and artificial intelligence have allowed large-scale collection and real-time analysis \cite{Yadav2023}. These advances extend OSINT applications across many fields. In cybersecurity, OSINT improves threat detection and vulnerability assessment \cite{Kovaci2024, Sidorova2023}. Law enforcement uses it for investigations, counterterrorism, and digital forensics \cite{Chaudhary2022, Elguindy2021}. It also supports national security by verifying sources and detecting leaks \cite{Yamin2022}, as well as conservation biology and the study of social or customer sentiment \cite{Katzner2022, Kruspe2024, Pai2021}.

Several researchers have proposed structured workflows for OSINT, which, despite minor variations, share a common structure. For example, \cite{Hwang2022} outlines a fundamental workflow that includes the following stages:
\begin{enumerate}
    \item Identify sources based on the intelligence objective
    \item Collect data from open and public sources
    \item Process data to clean and structure raw information
    \item Analyze data using techniques such as text mining, social network analysis, and geospatial mapping
    \item Report and disseminate findings
\end{enumerate}


The authors in \cite{Yadav2023} expand on the model of \cite{Akhgar2016} and stress the role of planning and dissemination, especially when the findings guide improvements in methods and techniques. \cite{Kumar2021} reports the same five phases, while \cite{Browne2024} highlights only the last four. \cite{PastorGalindo2020} focus on collection, analysis, and knowledge extraction, showing how OSINT techniques support data transfer during collection. The output from these steps generates useful insights for both individual and organizational analysis. 


Despite established workflows, several challenges remain. Large volumes of noisy data make filtering and extraction difficult \cite{Yadav2023, Hwang2022}. Automating these steps improves usability of the data \cite{Pelofske2023}, but integrating non-OSINT data requires careful attention to context and source reliability \cite{Day2016, Dokman2020}. OSINT resources often focus on specific regions, limiting global use \cite{PastorGalindo2020}. Tracking or de-anonymising individuals also presents difficulties \cite{Schafer2019}. Finally, rapid technological change forces OSINT methods to adapt continuously \cite{Szymoniak2024}.

\subsection{Peer-to-peer networks}

Peer-to-peer (P2P) networks follow a decentralized model, where each node acts as both a client and a server. Unlike client-server systems, which rely on central servers, P2P networks allow peers to share resources directly. This design improves scalability and fault tolerance \cite{Chien2009, Jaideep2016}, but also creates risks such as denial-of-service attacks, worm propagation, and content poisoning \cite{Jaideep2016}.



P2P networks can be structured or unstructured. Structured networks use predefined topologies and Distributed Hash Tables (DHT) for efficient resource discovery \cite{Mokadem2012}. Unstructured networks connect freely and rely on flooding to locate resources \cite{Jamal2017}. Although less efficient, unstructured networks adapt better to frequent peer turnover and work well in file-sharing systems like BitTorrent \cite{Shojafar2015}.


BitTorrent is the most widely used P2P file-sharing protocol, with over 100 million monthly users worldwide \cite{BitTorrent2023}. Although it solves the problem of distributing large files, it is also strongly linked to copyright violations: studies show that most shared content is unauthorized or illegal \cite{Watters2011, Schmidt2012}.  Two significant concerns related to BitTorrent usage are copyright infringement and the distribution of child exploitation material (CEM) \cite{Schmidt2012}. The lack of central control makes monitoring difficult and fuels legal disputes \cite{AlKhater2020}.


Another serious issue is the distribution of child exploitation material (CEM). Research shows that many offenders prefer BitTorrent over other platforms such as Gnutella or eMule \cite{Mutawa2015, Gautam2022}. The anonymity of P2P networks reduces the risk of detection and prosecution, making them attractive for criminal use \cite{AlKhater2020, Lee2020}. These networks attract offenders because they are free, open to all, and offer anonymity with little direct contact with providers \cite{Lee2020}.
\section{Related work}\label{chapter:relatedtwork}


Illegal activity on P2P networks is a persistent challenge that researchers study mainly in the context of copyright infringement and CEM distribution. Most work tracks content distribution, while a few studies focus on profiling individual users. While torrent metadata and traffic support research in cybersecurity, law enforcement and network optimization, their use for OSINT remains limited \cite{Riebe2023, Kovaci2024}.


Research on torrent data falls into three broad areas: (1) piracy and forensic investigations, (2) protocol optimization, and (3) network traffic analysis relevant to intelligence. Studies on piracy usually map activity on a national or regional scale. Although these studies provide valuable insights into P2P activity, they largely overlook the potential of torrent data for user profiling.


For example, the authors in \cite{Lareida2018} analyze video distribution on BitTorrent by country and evaluate antipiracy measures, while \cite{Kigerl2013} shows that smaller and poorer countries have relatively higher piracy rates. Although these studies map piracy trends and incorporate geographical data, they do not focus on individual users.


Torrent data also supports forensic investigations, particularly in cases related to CEM. The authors in \cite{Bissias2016} estimate the number of peers involved in five P2P networks and analyze trends in content distribution, but focus on file types rather than offender profiling. The authors in \cite{Mutawa2015} take a step toward profiling by applying behavioral evidence analysis (BEA) to P2P cases, showing how digital activity can reveal offender traits and support evidence reconstruction. However, their work focuses on individual cases, rather than large-scale profiling, and does not incorporate open data, such as IP metadata, leaving a gap in the research.


A large body of work improves the BitTorrent protocol itself. Studies optimize file transfers, reduce cross-ISP traffic, and design better routing and tracking mechanisms \cite{Yu2010, Li2019, Gartner2023, Balkibayeva2024, Barreto2024, Lajam2024, Hwang2022a, Nicolini2019, Simion2019}. Other work examines the popularity of content and network usage to improve bandwidth provisioning \cite{Scanlon2014, Lareida2017} or recommendation systems \cite{Han2019}. While valuable for network performance and efficiency, these studies do not address open source intelligence gathering.

Closer to OSINT, traffic analysis provides useful methods. In \cite{Sainani2020}, the authors build an anomaly detection model that scores IP addresses based on behavior and location, while \cite{Bagui2017} and \cite{Lotfollahi2019} classify encrypted traffic and VPN usage for network security purposes. These approaches improve cybersecurity, but not user profiling. More relevant examples come from deanonymization and behavioral clustering. In \cite{Torres2024}, it is shown that user interests can be reconstructed from anonymized rating datasets, suggesting similar techniques for torrent downloads. \cite{Mateless2021} cluster entities behind a shared IP based on browsing traits, supporting behavioral attribution. These findings align with the goals of OSINT profiling, as torrent downloads can reveal patterns of interest associated with individual users. \cite{Ren2022} further highlights the challenge of linking users to IP addresses using multisource data, such as certificates and banners. However, their work remains limited in scope and constrained to specific datasets and regions.


Despite substantial research on torrent activity and network behavior, the literature reveals several important gaps that this work aims to address.
\begin{itemize}
\item \textbf{Lack of user-level profiling:} Most research focus on trends, networks, or protocol efficiency, not individual behavior \cite{Balkibayeva2024, Gartner2023, Nicolini2019, Simion2019, Lareida2018, Bissias2016}.
\item \textbf{OSINT methods are underused:} Although common in cybersecurity and social media analysis \cite{PastorGalindo2020, Yadav2023}, OSINT rarely extends to P2P data.
\item \textbf{Cross-referencing public data is limited:} Forensic studies \cite{Mutawa2015} link behavior to evidence but do not integrate open sources.
\item \textbf{Focus on content rather than context:} Research often tracks file categories or popularity instead of user behavior across content \cite{Scanlon2014, Han2019}.
\end{itemize}


To address these gaps, we present a proof-of-concept that applies OSINT methods to publicly accessible P2P data. The workflow combines torrent metadata, tracker responses, and IP enrichment to support behavioral profiling of users engaged in high-risk activity. This approach highlights the largely unexplored intersection between torrent data and structured OSINT profiling, opening a new direction for intelligence research.

\section{Methodology}\label{chapter:methodology}

\begin{figure}[h]
    \centering
    \makebox[\textwidth]{\includegraphics[scale=0.33]{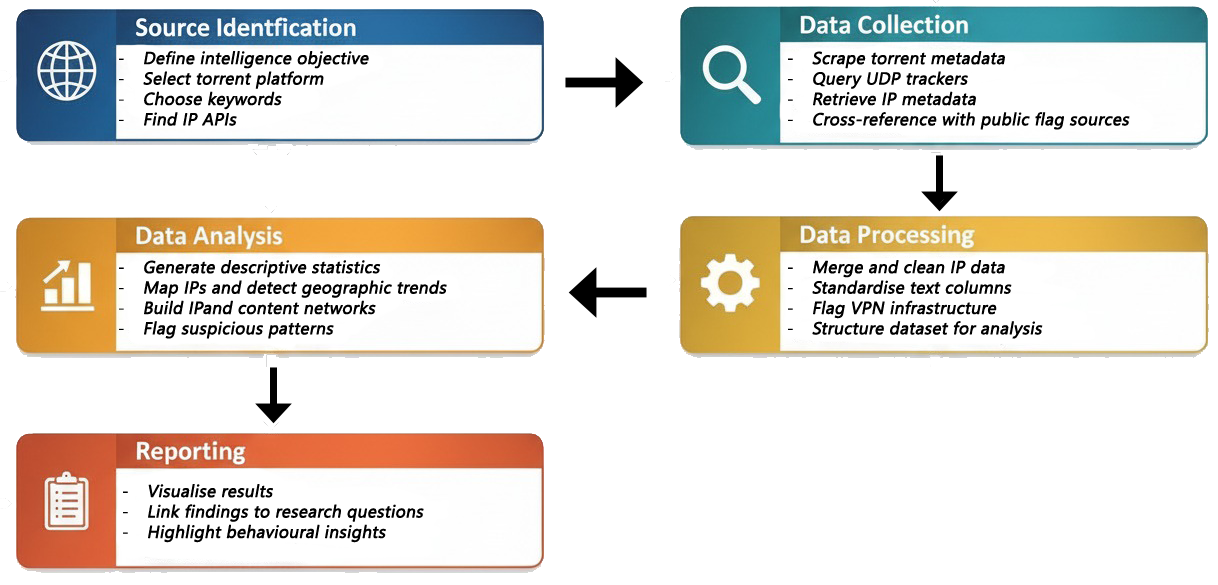}}
    \caption{Data pipeline design} 
    \label{fig:data_pipeline_design}
\end{figure}

The workflow we propose follows the structure outlined in \cite{Hwang2022} and applies OSINT methods to profile and de-anonymize BitTorrent users. It uses a data pipeline that integrates multiple sources, processes collected information, and extracts insights through systematic analysis. The pipeline consists of five stages: source identification, data collection, data processing, data analysis, and reporting (Figure \ref{fig:data_pipeline_design}). Each stage connects to the next, supporting a comprehensive and iterative investigation.

The following sections provide a detailed description of each stage, with an emphasis on the sources, techniques, and tools employed.

\subsection{Source identification}\label{section5.1}

OSINT becomes powerful when it combines different open data sources with analytical methods to build continuous profiles of targets \cite{PastorGalindo2020}. With this approach, we designed a study to profile and de-anonymize BitTorrent users, especially those downloading flagged content. We focused on four main types of open-source data: (1) torrent index sites for file and uploader metadata, (2) tracker responses to identify active peers, (3) IP metadata from public geolocation and infrastructure services, and (4) torrent monitoring databases that flag high-risk behavior. Together, these datasets create a multi-angle view of user activity and patterns.



BitTorrent revolves around trackers and magnet links. A tracker is a central server that coordinates swarms (the term swarm refers to the entire network of peers) by sharing lists of peers that exchange a specific file \cite{Wu2011}. More peers in a swarm generally lead to faster downloads \cite{Yu2010, Scanlon2014}. Alongside trackers, Distributed Hash Tables (DHT) enable peers to discover each other without relying on central servers, making the network more resilient \cite{Schutz2021}. Magnet links push resilience further by replacing .torrent files with cryptographic hash values. This eliminates the need for a fixed location and enables peers to connect directly through the DHT or public trackers \cite{Jarosz2022, Schutz2021}.


While efficient, BitTorrent exposes privacy risks because its peer-to-peer structure makes IP addresses visible to anyone in the swarm \cite{Simion2019}. Unlike client-server downloads, where only the server sees requests, peers in BitTorrent can watch each other’s activity. Monitoring groups, copyright enforcement agencies, or malicious actors can easily exploit this. Although some users rely on VPNs or proxies, many remain exposed due to BitTorrent’s design.


For data collection, we used four sources. First, we gathered metadata from torrent index sites. We selected \textit{The Pirate Bay}, a popular public index with a high number of active torrents, uploaders, and trackers \cite{Martin2016, Zhang2011}. From its Top 100 Torrents page, we extracted torrent names, categories, upload dates, file sizes, seed and leech counts, uploader information, and magnet links. Seeders hold the full file and share it, while leechers only download (and do not share). Seeder and leecher counts reveal both the popularity and the health of a torrent as more seeders generally lead to faster downloading speeds. This dataset highlights file trends, but deeper insights emerge when it is combined with other sources.


To study how torrent data can reveal high-risk or suspicious behavior, we ran keyword searches on The Pirate Bay. We searched for ``explosive(s)'', ``IED'', and ``bomb(s)'', as these terms often appear in media and law enforcement discussions about improvised explosive devices\footnote{While not drawn from a formal keyword list, these terms are widely associated with improvised explosive devices in public, academic and investigative contexts \cite{Kumar2020, Ferreri2019}.}. Each search returned multiple torrents from a single uploader, \textbf{crwildman}, who appeared across all results. His uploads under ``IED'' and ``bomb(s)'' keywords included guides on improvised explosives, bombmaking, and homemade chemical compounds. Due to his recurring presence and the large volume of relevant material he uploaded in 2013, we selected the 59 books he uploaded as a representative subset. This dataset allowed us to analyze swarm activity and user behavior around a clearly defined cluster of high-risk content.


The second data source is tracker response data. We scraped UDP trackers linked to torrents to identify active peers. While DHT scraping can reach a broader and more resilient set of users, we chose central UDP trackers due to technical limitations. Using magnet links from The Pirate Bay, we retrieved file hashes and announced them to the trackers. The trackers then returned IP addresses and ports of connected peers, giving us the real-time swarm distribution and its composition.


The third source was IP metadata. We utilized public APIs (ip-api.com and ipinfo.io) to enrich IP addresses with geolocation, ISP, organizational ownership, and autonomous system (AS) numbers, which are unique identifiers assigned to groups of IP prefixes managed by a single organization, offering additional insight into infrastructure ownership and hosting patterns. These services also provided indicators on anonymity measures. The metadata helped us map the geographic spread, detect hosting providers, and identify users without strong anonymity measures.


Finally, we used an external monitoring database (iknowwhatyoudownload.com). It collects publicly visible torrent activity linked to IPs and labels them with behavioral tags. The database flags IPs that have previously downloaded sensitive or illegal content, including CEM. We did not store or process such material ourselves due to ethical and legal constraints. Instead, we only cross-referenced our collected IPs with those flagged entries. This enhanced our risk profiling without requiring us to handle illicit data directly.

\begin{figure}[!h]
    \centering
    \begin{adjustbox}{center}
        \includegraphics[scale=0.6]{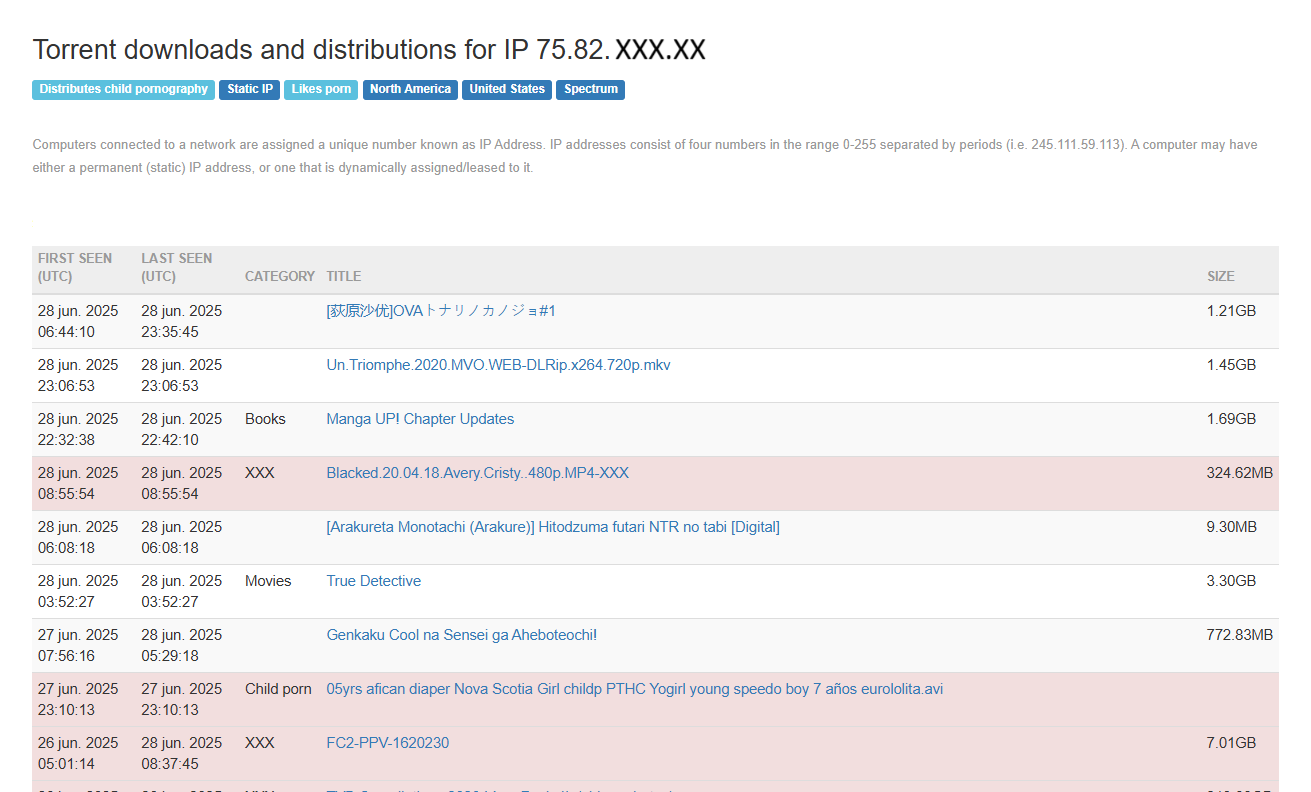}
    \end{adjustbox}
    \caption{Screenshot of \textit{iknowwhatyoudownload.com}} 
    \label{fig:iknowwhat}
\end{figure}

\subsection{Data collection}


We combined multiple tools and libraries to scrape data from different web pages. We employ Selenium WebDriver, an open-source framework for automating web browsers and simulating user interactions. This approach enables us to bypass common restrictions on automated scraping and extract website content. To handle rate limits and reduce detection, we also added delays between requests. We parsed the extracted HTML using BeautifulSoup and exported the relevant data into a CSV file for further analysis.



For UDP scraping, we built a custom socket-based scraper by adapting the m2t (magnet-to-torrent) project \cite{m2t}. While the original m2t focused on scrape requests for simple metadata, we had already gathered this through web scraping from The Pirate Bay. Instead, we modified the code to send announce requests, which return real-time peer lists for active swarms. Announce requests query the tracker for real-time information about peers currently participating in a torrent swarm, which the original implementation did not capture. This required restructuring request payloads and rewriting parsing logic to interpret binary tracker responses.


UDP trackers use binary byte buffers (packets) defined by the BitTorrent protocol \cite{BEP15}. Our process begins with a connection request to the tracker containing a connection ID and transaction ID. The tracker then issues a new connection ID, which we use to send announce requests. These do not download the torrent but instead query the tracker for currently active peers. The tracker’s response encodes each peer in six bytes: the first four for the IP address and the last two for the TCP port. This method enables efficient data extraction from UDP trackers, providing real-time insights into the swarm composition and active participants.


The pipeline proved both functional and productive. In total, we scraped metadata for 206 unique torrents, collected 60,369 unique IP addresses from tracker responses, and enriched them with metadata from public IP sources. We also gathered 42 IPs from torrents involving sensitive books (see Section \ref{section5.1}) and cross-referenced all collected IPs with public flagging sources, identifying 940 addresses previously linked to CEM distribution. Complete descriptive statistics and dataset breakdown appear in Section \ref{:data}.

\subsection{Data processing}


After collecting IP addresses and enriching them with metadata, we built a dataset with one record per IP. Each record included city, country, ISP, AS number, latitude and longitude, resolved hostname, anonymity and anycast indicators, and a flag for known involvement in CEM distribution. Anycast refers to IPs shared by multiple systems across multiple locations. We combined and cleaned these records to ensure consistency.


We merged CSV files from batch collections into one dataset, removed duplicates, and standardized text fields. ISP and organization names were lowercased, and inconsistent country and city entries were corrected. City names were cross-referenced with regional data to prevent false merges. For example, ``Dar Es Salaam'' was unified as ``Dar es Salaam'', while distinct locations with similar names, such as ``Lafayette'' (Louisiana) and ``LaFayette'' (Georgia), remained separate. These corrections improved the accuracy of aggregation, although for geospatial mapping, latitude and longitude alone would be sufficient.


The dataset also contained two boolean fields: Anycast and Privacy. Since no IPs were flagged as Anycast, we dropped this field. Privacy indicates whether an IP relates to a VPN, proxy, or hosting provider. The Privacy field indicates whether the IP is associated with a VPN, proxy, or hosting provider. To capture missing cases not flagged by \textit{ipinfo.io}, we scanned ISP and AS fields for keywords such as ``Cloud'', ``Host(ing)'', ``Data Centre/Center'', ``Private Internet'', ``Private network'', ``Proxy'', ``TOR'', and ``VPN''. If present, we set the Privacy flag to True.


Through this process, we created a standardized and enriched dataset, ready for downstream analysis such as clustering by geography, ISP, or anonymity. These preparations laid the groundwork for detecting outliers and identifying patterns associated with high-risk torrent activity.

\subsection{Data analysis}
After preprocessing, the cleaned and structured dataset is analyzed to identify patterns in torrent participation, geographic distribution, privacy measures, and flagged activity. This analysis combines statistical methods and data visualization to reveal key trends.
\begin{itemize}
    \item Descriptive statistics: We summarize key attributes such as torrent categories, geographic IP distribution, use of anonymity measures, and content popularity to identify general patterns in user and file characteristics.
    \item IP network analysis: IP-to-IP graphs based on shared downloads reveal relationships among users and visualize swarm-level interaction patterns.
    \item Content network analysis: We construct graphs linking torrents based on co-downloads to explore thematic clusters and cross-topic interests.
    \item Outlier detection: High-activity or focused-interest IP addresses are flagged as potential outliers and profiled to assess risk relevance.
    \item Anonymization detection: We assess the prevalence of VPNs, proxies, and cloud-hosted IPs, correlating these with sensitive content downloads to evaluate anonymity strategies.
    \item Visualization and reporting: Graphs, maps, and charts support the interpretation of results and communicate key findings effectively.
\end{itemize}

\subsection{Report and disseminate findings}


The final stage of our workflow structures, interprets, and presents the results in a form that enables understanding and actionable insights. We combine network visualizations, geospatial mapping, statistical summaries, and categorical breakdowns to highlight patterns in user behavior, content relationships, and risk indicators. These outputs are designed for OSINT analysts, researchers, and investigators.


We use three main types of visualizations: (1) unipartite network graphs to show links between IP addresses, content, and co-download behavior; (2) geospatial heat maps and dot maps to display the global distribution of torrent participants and clustering of sensitive activity; and (3) bar charts and tables to summarize torrent categories, IP features, anonymity use, and download frequencies. Each visual includes a narrative explanation to connect the results directly to our research questions.



This reporting approach transforms raw metadata into intelligence signals that improve the interpretability of OSINT methods applied to torrent ecosystems. Beyond supporting this study, the visual and analytical framework also refines wider OSINT practices. By sharing these results within the research community, we contribute to advancing techniques for analyzing peer-to-peer network activity, online anonymity, and digital investigation more broadly.

\section{Data and descriptive statistics}\label{:data}


We stored and analyzed the collected data using a relational database with three interconnected tables: ip\_info, tor\_info, and hash\_ip. These tables systematically link IP addresses, torrent metadata, and shared files. 



The ip\_info table uses IP addresses as unique identifiers and contains location, ownership, and privacy attributes. Each record includes city, regional subdivision, country, Internet Service Provider, organization, and Autonomous System number. We also stored latitude and longitude coordinates, resolved hostnames, privacy service flags, and flags for IPs linked to CEM distribution.


The tor\_info table holds torrent metadata with file hashes as unique identifiers. Attributes include title, category, subcategory, upload date, file size, seeder and leecher counts, uploader username, and magnet link. We classified books uploaded by \textbf{crwildman} into six interest categories: \textit{government and military operations}, (2) \textit{assassination, killing, and combat techniques}, (3) \textit{DIY explosives, weapons, and drugs}, (4) \textit{survival and evasion techniques}, (5) \textit{lockpicking, pickpocketing, and safe manipulation}, and (6) \textit{cannibalism}. 


The hash\_ip table connects torrents to corresponding IP addresses, linking users with shared files. It contains hash values referencing tor\_info records and IP addresses linking to ip\_info entries.

\begin{figure}[h!]
    \centering
    \includegraphics[scale=0.15]{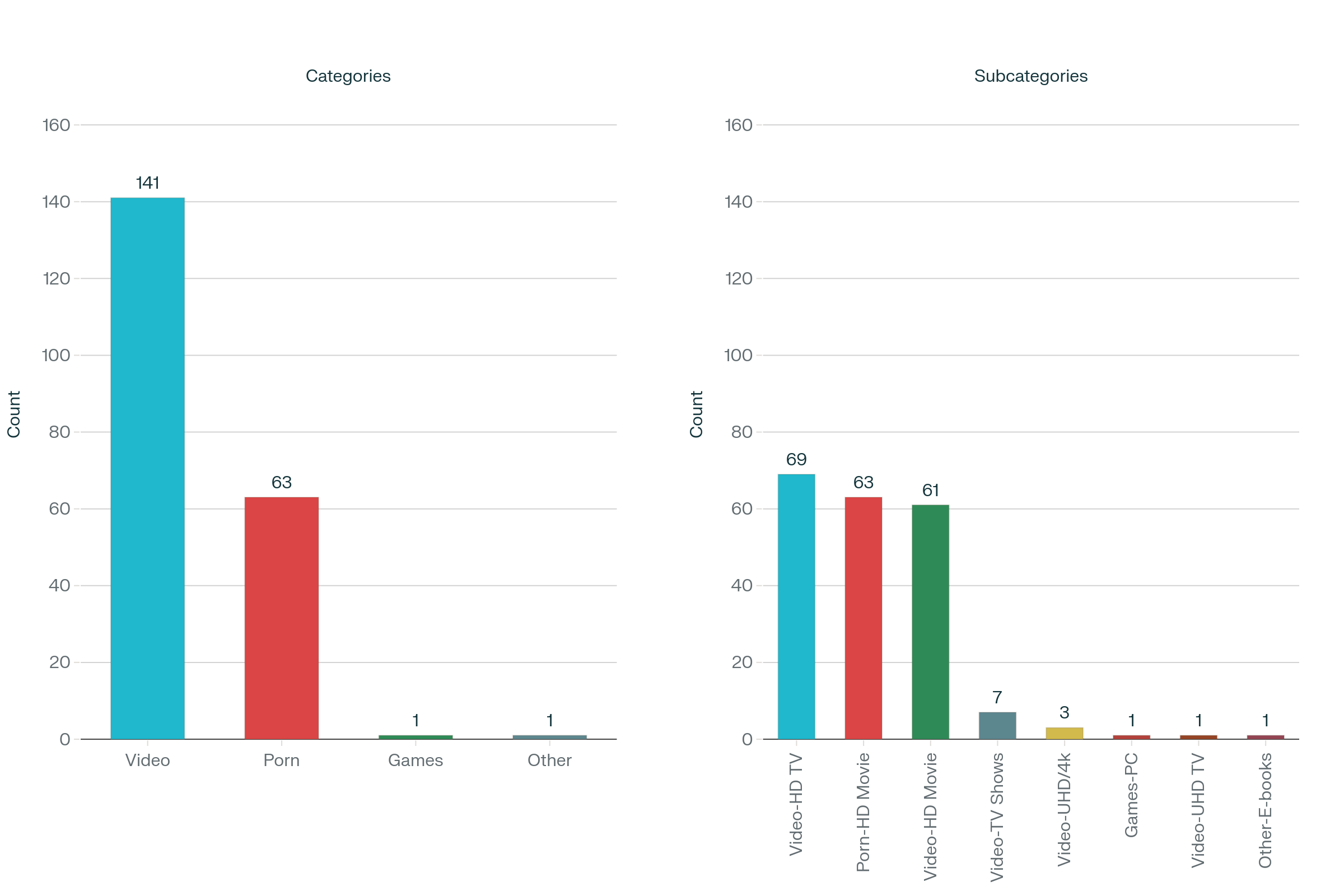}
    \caption{Counts per category}
    \label{fig:category_counts}
\end{figure}

We collected data by scraping the \textit{Top 100 Torrents} page daily from February 8th to March 4th, 2024, yielding 206 unique torrents. The dataset included 141 TV and film torrents, 63 pornographic titles, one game, and one e-book. Figure \ref{fig:category_counts} provides a visual breakdown of the dataset by category and subcategory, offering a clearer picture of the relative proportions and dataset size. Only 9 uploaders contributed these 206 torrents, with 3 users uploading just one title each. 

\begin{figure}[h!]
    \centering
    \includegraphics[scale=0.17]{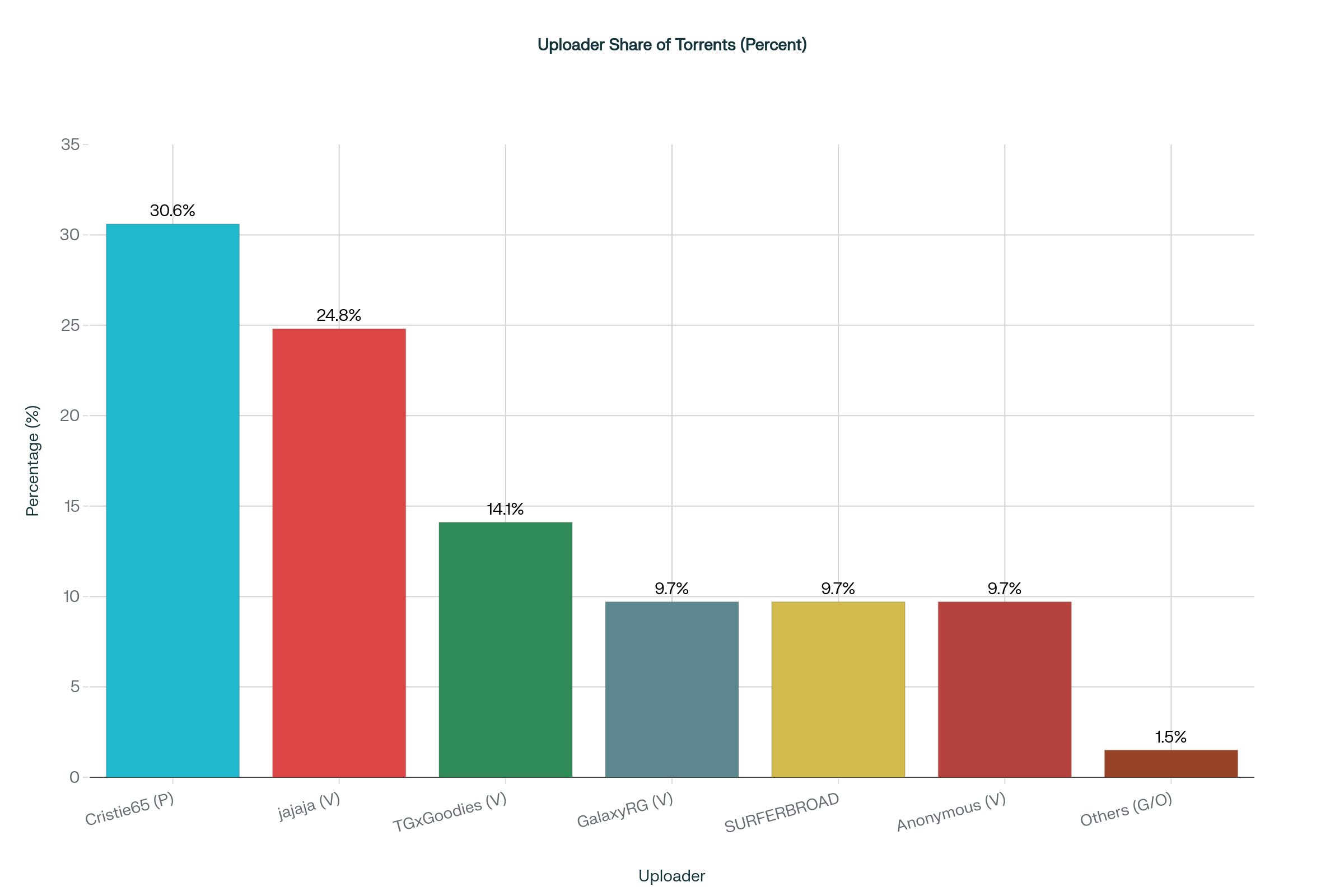}
    \caption[Uploaders of popular torrents]{Uploaders of popular torrents and their percentage share (of the 206 torrents)}
    \label{fig:uploaders}
\end{figure}

Figure \ref{fig:uploaders} displays the distribution, revealing that \textit{Cristie65} uploaded all pornographic content during our collection period.

Table \ref{tab:summary_stats} summarizes key statistics for both the popular torrents and the flagged books. Each metric is discussed in more detail in the following paragraphs. 


\begin{table}[t]
\centering
\scalebox{0.8}{
\begin{tabular}{p{0.45\textwidth} p{0.28\textwidth} p{0.43\textwidth}}
\toprule
\textbf{Metric} & \textbf{Popular Torrents} & \textbf{Flagged Books (Case Study)} \\
\midrule
Number of torrents & 206 & 59 \\
Unique IP addresses & 60.369 & 42 \\
Countries represented & 213 & 19 \\
Flagged for CEM & 940 (1,56\%) & 3 (7,14\%) \\
Flagged for privacy (VPN/proxy) & 12.392 (20,53\%) & 15 (35,71\%) \\
\bottomrule
\end{tabular}
}
\caption{Summary statistics for torrent datasets}
\label{tab:summary_stats}
\end{table}




The thematic patterns in the popular torrents dataset include ``\textit{sex}'', ``\textit{body}'', ``\textit{sister}'', and ``\textit{female}'', pointing to a strong representation of pornographic or adult-themed content. Simultaneously, terms like ``\textit{detective}'', ``\textit{halo}'', ``\textit{airbender}'', and ``\textit{avatar}'' reflect continued interest in mainstream entertainment genres such as crime, science fiction, and action. These findings support earlier quantitative results (Figure \ref{fig:category_counts}), which show that the majority of torrents fall into either film/TV or pornography categories.



We scraped associated trackers and obtained 60,369 unique IP addresses from 213 countries. Each IP address is connected to an average of 1.51 downloads of popular torrents, with a maximum of 57 downloads. Our analysis revealed that 940 IPs (1.56\%) were flagged for known CEM distribution. Furthermore, we identified 12,392 IP addresses (20.53\%) as highly likely users of privacy-preserving services such as VPNs or proxies. Among the remaining IPs, 16,591 addresses (27.48\%) showed neither privacy flags nor identifiable hostnames, making them the least anonymized group. This non-anonymized subset contained 79 CEM-flagged IP addresses, representing 8.40\% of all CEM-related addresses. In contrast, 75.85\% of CEM-flagged addresses used privacy-preserving technologies, indicating significantly higher anonymization rates among users involved in this activity. This pattern suggests that individuals downloading sensitive content actively employ privacy tools to conceal their identities. While most IP metadata was complete, many entries lacked organizational data, limiting detailed analysis options. However, we successfully captured ISP information for all IPs. Table \ref{tab:isp} shows the ten most commonly used ISPs in our dataset.


\begin{table}[h!]
\centering
\scalebox{0.8}{
\begin{tabular}{p{0.5\textwidth} p{0.3\textwidth} p{0.3\textwidth}}
\toprule
\textbf{ISP} & \textbf{Captured IP count} & \textbf{Percentage of total} \\
\midrule
datacamp limited & 1486 & 2,46\% \\
tefincom s.a. & 1315 & 2,18\% \\
m247 europe srl & 792 & 1,31\% \\
philippine long distance telephone co. & 785 & 1,30\% \\
chinanet & 768 & 1,27\% \\
claro nxt telecomunicacoes ltda & 694 & 1,15\% \\
cogent communications & 615 & 1,02\% \\
clouvider limited & 489 & 0,81\% \\
rcs \& rds & 473 & 0,78\% \\
packethub s.a. & 462 & 0,77\% \\
\bottomrule
\end{tabular}
}
\caption[Most common internet service providers]{Most common internet service providers and their share of total IPs ($n = 60{.}369$)}
\label{tab:isp}
\end{table}


The most represented ISP is \textit{Datacamp Limited}, accounting for 2.46\% of the 60,369 captured IP addresses. Nearly 99\% of these IPs use privacy-preserving technologies, which aligns with the company's role as a major infrastructure provider for commercial VPN services such as Surfshark and NordVPN.


Anonymization correlates strongly with higher-risk signals. While 1.56\% of all IPs carry CEM flags, this rate increases to 7.81\% among IPs associated with \textit{Datacamp Limited}. This pattern confirms our earlier finding that users who engage in high-risk activities are more likely to adopt anonymization strategies.


\textit{Tefincom S.A.} accounts for 2.18\% of IP addresses and has historical ties to NordVPN development. The data reflects this connection: 100\% of \textit{Tefincom} IPs show VPN or proxy flags.



Nine of the ten most common ISPs display similarly high anonymization rates, with over 95\% of their IPs flagged for privacy-preserving technologies. This trend highlights the widespread use of VPN infrastructure in torrent ecosystems. The sole exception is \textit{Claro NXT Telecomunicações Ltda}, a major Brazilian ISP with no anonymization flags, suggesting more direct, non-obfuscated user activity.

\begin{figure}[h!]
    \centering
    \includegraphics[scale=0.33]{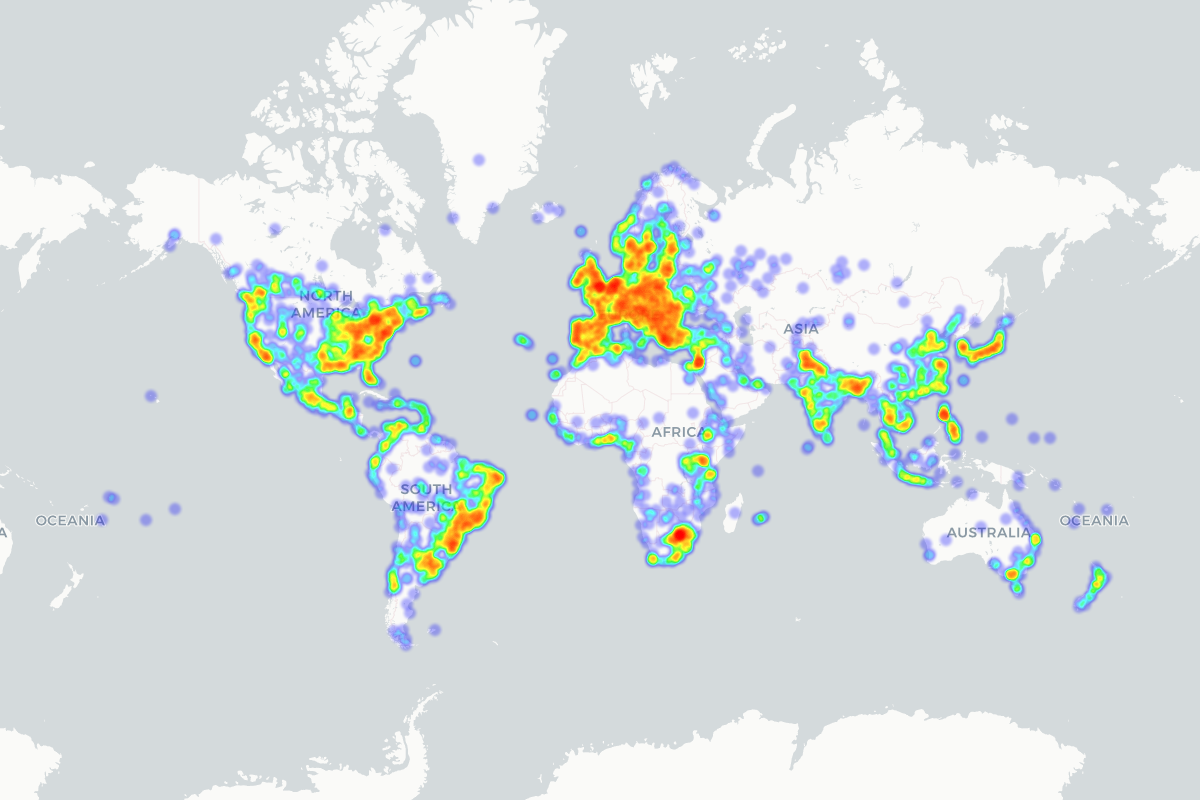}
    \caption{Geographical distribution of popular torrent activity}
    \label{fig:ip_distribution}
\end{figure}

We plotted geolocation data in a heat map for spatial analysis. Figure \ref{fig:ip_distribution} shows the global distribution of IP addresses by country, with the United States, United Kingdom, Canada, Brazil, and South Africa most represented. This distribution aligns with \cite{Kigerl2013}, who found that wealthier countries exhibit higher absolute piracy levels. Since heat maps emphasize patterns rather than exact counts, Table \ref{tab:top_countries} complements the visualization by listing the ten countries with the most IP addresses and their dataset percentages.

\begin{table}[h!]
\centering
\scalebox{0.8}{
\begin{tabular}{p{0.4\textwidth} p{0.12\textwidth} p{0.12\textwidth}}
\toprule
\textbf{Country} & \textbf{Count} & \textbf{Share} \\
\midrule
United States of America & 5.181 & 8,59\% \\
United Kingdom & 3.263 & 5,41\% \\
Canada & 3.037 & 5,03\% \\
Brazil & 2.835 & 4,70\% \\
South Africa & 2.733 & 4,53\% \\
Australia & 2.198 & 3,64\% \\
Philippines & 1.871 & 3,10\% \\
Netherlands & 1.815 & 3,01\% \\
China & 1.768 & 2,93\% \\
Portugal & 1.528 & 2,53\% \\
\bottomrule
\end{tabular}
}
\caption[Top 10 countries by number of IP addresses]{Top 10 countries by number of IP addresses in the dataset ($n = 60{.}369$)}
\label{tab:top_countries}
\end{table}


To examine suspicious downloading behavior, we analyze the 59 book titles that user `\textit{crwildman}' uploaded in January 2013. While 23 of these torrents currently show no active participants and only 15 have multiple peers, these files still attract download activity over a decade later.



\begin{figure}[h!]
    \centering
    \includegraphics[scale=0.17]{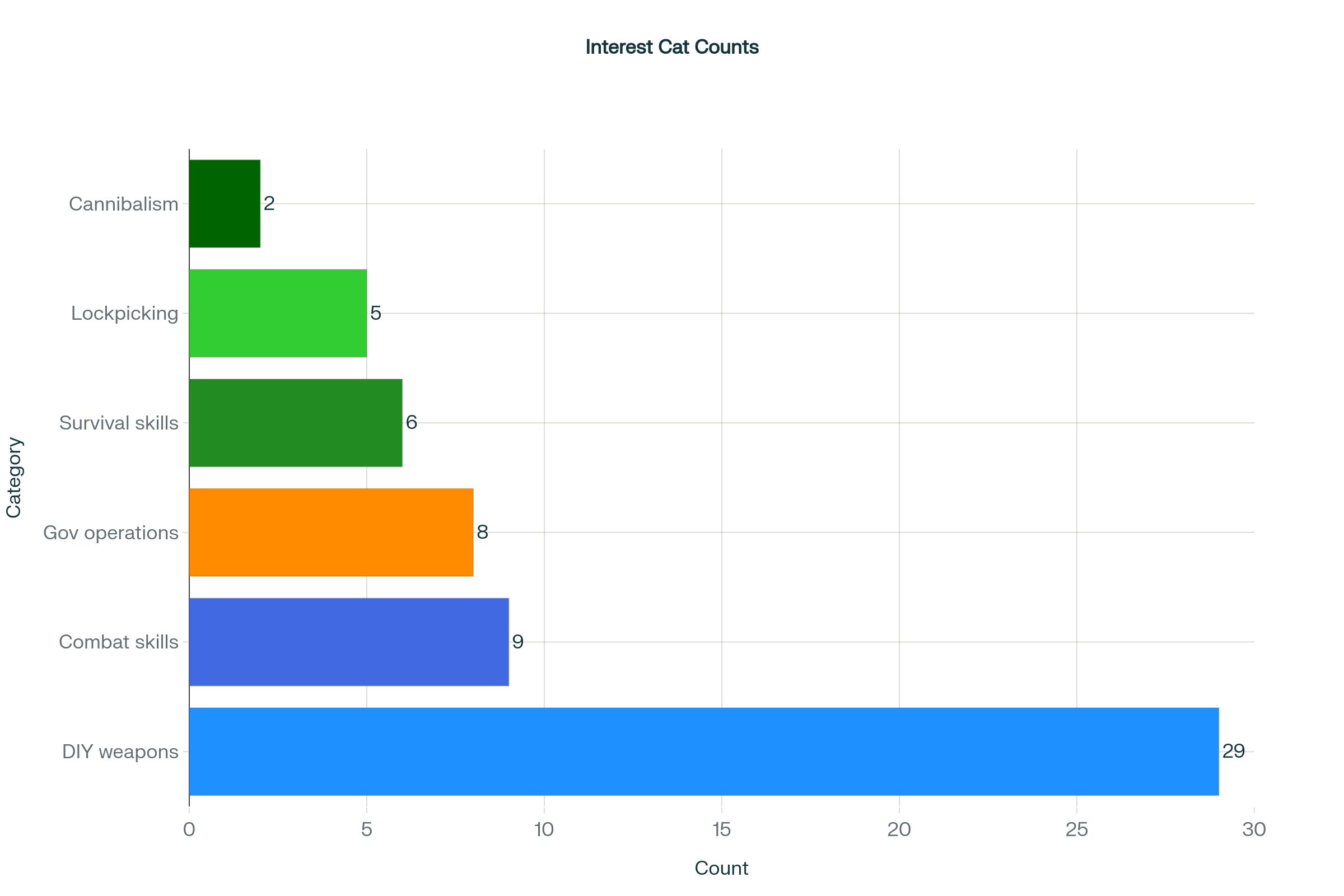}
    \caption{Book counts per category}
    \label{fig:interest_category_count}
\end{figure}

Figure \ref{fig:interest_category_count} shows the distribution of book titles across content categories. The `\textit{DIY explosives, weapons, and drugs}' category contains the most books and accounts for the majority of active downloads. The three torrents with the most active peers all relate to explosives: \textit{MILITARY Explosives Chemistry Must Have Ebook}, \textit{A Guide to Field Manufactured Explosives.pdf}, and \textit{Explosives and Weapons - Homebuilt Claymore Mines}. We found four active peers for each torrent during our measurements.

Among the 59 books uploaded by the user `\textit{crwildman}', the most frequently occurring terms include ``\textit{explosives}'', ``\textit{manual}'', ``\textit{survival}'', ``\textit{interrogation}'', ``\textit{attack}'', and ``\textit{CIA}'', reflecting a strong orientation toward operational tactics, improvised weaponry, and intelligence-related techniques. Other terms such as ``\textit{guide}'', ``\textit{hypnosis}'', ``\textit{poison}'', and ``\textit{guerrilla}'' indicate the presence of psychological and chemical warfare content, along with survivalist and subversive tactics.



Compared to the general torrent dataset, this material exhibits significantly higher levels of violent or illegal instructional content. The content focus matches our categorization results and supports treating this collection as an important risk factor in OSINT profiling (due to limited space, we do not provide the complete details of the 59 book titles, including their categories and peer counts, for each torrent here).

\begin{figure}[h!]
    \centering
    \includegraphics[scale=0.35]{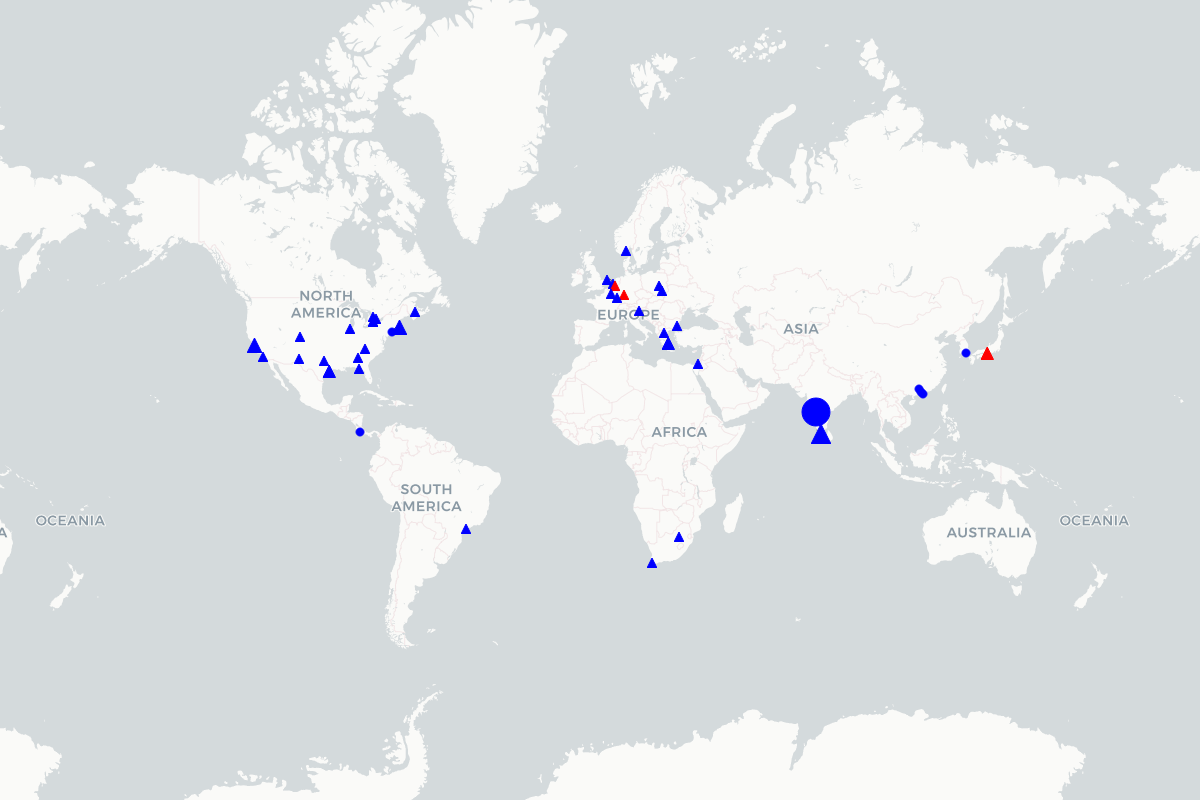}
    \caption{Geographical distribution of IP addresses downloading sensitive books}
    \label{fig:crwildman_geo_distribution}
\end{figure}

We scraped the e-book swarms and identified 42 unique IP addresses, with two of them also appearing in the popular torrent dataset. These IPs come from 19 countries: 12 from the United States, 5 from Canada, 3 each from China and the Netherlands, and 1-2 from other countries. Figure \ref{fig:crwildman_geo_distribution} illustrates the distribution of this location, where larger shapes indicate a higher number of downloaded books. Circles indicate IP addresses least likely to use anonymization, while red markers show IPs associated with CEM distribution.

We found that three of the 42 IP addresses carry CEM distribution flags. Two of these accessed DIY explosives material: one downloaded a guide to field-manufactured explosives, while the other got chemical supplier lists. The third IP is associated with a book on combat technique content.


Approximately one-third of the IP addresses are likely to use anonymization through VPN or proxy services. Among the seven addresses least likely to use anonymization, most accessed DIY explosives books.


Table \ref{tab:top_countries_sensitive_books} shows the key data from Figure \ref{fig:crwildman_geo_distribution}. For countries with multiple IP addresses, the table lists total downloads, IP counts, usage of privacy services, and CEM-flagged IPs. While the map provides an intuitive spatial overview, the table offers a clearer and more accessible breakdown of the underlying data.


\begin{table}[h!]
\centering
\scalebox{0.7}{
\begin{tabular}{p{0.18\textwidth} p{0.15\textwidth} p{0.22\textwidth} p{0.25\textwidth} p{0.29\textwidth}}
\toprule
\textbf{Country} & \textbf{Downloads} & \textbf{Number of IPs} & \textbf{VPN/Proxy Use} & \textbf{CEM Flagged IPs} \\
\midrule
United States & 17 & 12 & 7 & 0 \\
India & 14 & 2 & 0 & 0 \\
Canada & 5 & 5 & 3 & 0 \\
China & 3 & 3 & 0 & 0 \\
Netherlands & 3 & 3 & 2 & 1 \\
Greece & 3 & 2 & 0 & 0 \\
Poland & 2 & 2 & 1 & 0 \\
South Africa & 2 & 2 & 0 & 0 \\
\bottomrule
\end{tabular}
}
\caption{Top countries by number of downloads of flagged books}
\label{tab:top_countries_sensitive_books}
\end{table}

The United States stands out with the highest number of downloads (17) and unique IPs (12). Users from Canada and the Netherlands exhibit relatively high VPN or proxy use rates, indicating a higher level of privacy awareness.

One striking observation is the activity linked to India, where only two IP addresses account for 14 downloads. In Figure \ref{fig:crwildman_geo_distribution}, this is visible as two disproportionally large dots, with the largest representing a single IP address responsible for nine downloads. This concentrated activity suggests a targeted interest in the content and will be examined more closely in the following sections.
\section{Experimental Results}\label{chapter:experimentalresults}


To test how well our method can map torrent user networks using public data, we built a bipartite graph with two node types: IP addresses and torrents. We then converted this into a unipartite IP-to-IP graph, where edges indicate co-participation—i.e., two IPs appearing in the same swarm at the time of scraping.


Since the large number of IPs made the network hard to interpret, we filtered to include only the top 0.01\% of IP-to-IP connections by frequency. These edges represent the strongest behavioral links—repeated co-downloads across different torrents that likely indicate consistent user behavior patterns or shared networks.
We set a high activity threshold for network inclusion: IPs must have downloaded at least seven popular torrents. Among the qualifying IPs, we found clear content preferences: 217 downloaded only TV and film content, 140 focused exclusively on pornography, 11 accessed both categories, and 4 downloaded books and TV content.

\begin{figure}[h!]
    \centering
    \includegraphics[scale=0.35]{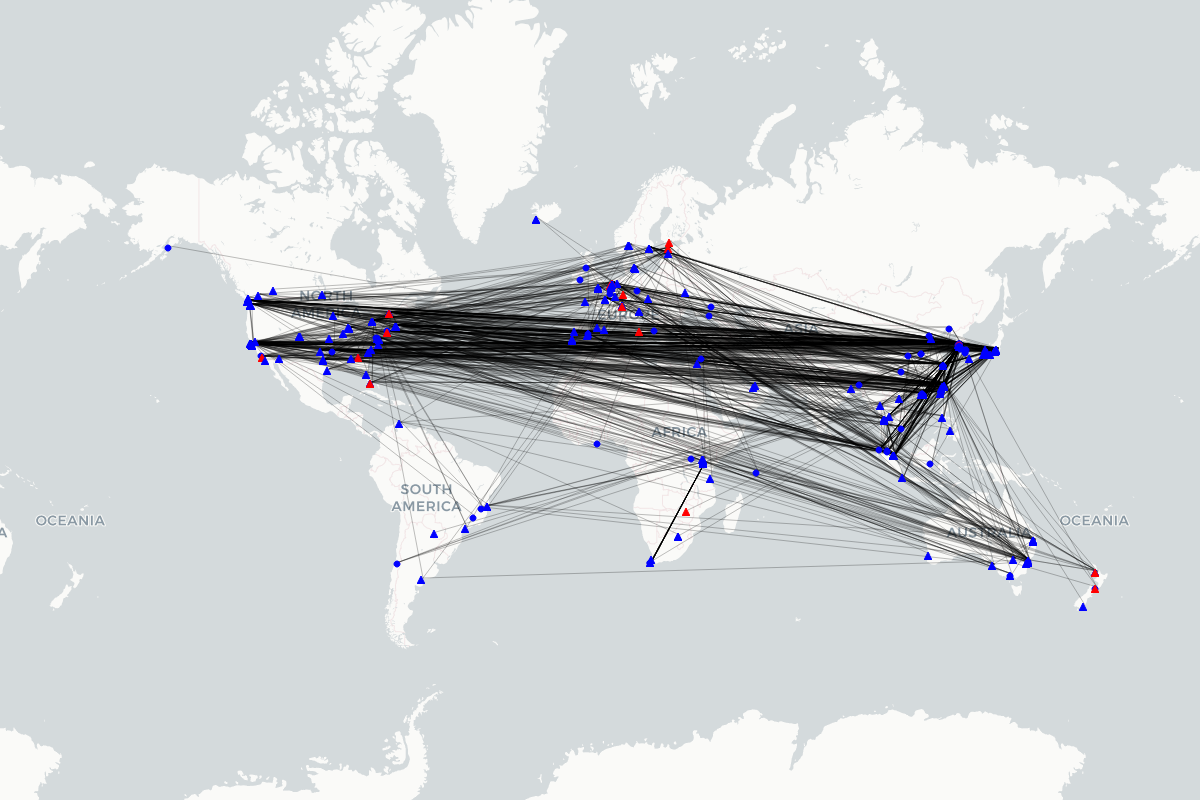}
    \caption{Network with 0,01\% connection frequencies.}
    \label{fig:networktop}
\end{figure}

Figure \ref{fig:networktop} visualizes this high-frequency network, comprising 372 IP addresses. Each node is geolocated based on IP metadata, and node styling indicates both anonymity and CEM-flag status: 
\begin{itemize}
    \item Circle = not flagged for anonymization
    \item Triangle = anonymized IP (VPN/proxy detected)
    \item Red = flagged for known CEM distribution
    \item Blue = non-flagged
\end{itemize}

The network reveals a globally distributed user base with several dense clusters. Geographic clustering is evident throughout, with many high-frequency connections occurring within specific regions. However, intercontinental ties also exist frequently, highlighting the decentralized nature of torrent swarms and the international reach of content on \textit{The Pirate Bay}.


Edge thickness encodes co-download frequency, with thicker lines showing more frequent shared swarm participation. However, these visual differences remain subtle at this level of filtering. To better show the most significant relationships, Table \ref{tab:top_ip_pairs} 
lists the top ten most frequent IP-to-IP co-download pairs. Each row captures a unique IP pair that appeared repeatedly in the same torrent swarms, suggesting a strong behavioral link through consistent co-download timing, shared interests, or possibly shared infrastructure.



For privacy and ethical reasons, the table displays approximate coordinates and countries for each IP address rather than the raw addresses. These serve as pseudonymous identifiers, allowing tracking across network figures and tables without exposing private information. The table also displays the anonymization status and whether either address has CEM distribution flags.
All high-frequency interactions in this subset involved pornographic content. Several IP addresses (marked by cell color) appear in multiple pairings, suggesting either highly active users or nodes acting as persistent peers across multiple swarms—systematic behavior worth further investigation.

\begin{table}[h!]
\caption[Top 10 most frequent IP (coordinate) pairs]{Top 10 most frequent IP (coordinate) pairs}
\captionsetup{format=plain}
\caption*{\small\textit{Note:} Cell colors indicate recurring IP addresses based on coordinates. Identical colors reflect the same IP across pairings. These colors are consistent with IPs shown in Table \ref{tab:cem_centrality_extended}
}
\resizebox{\textwidth}{!}{
\begin{tabular}{llrlrrrr}
\toprule
\textbf{Coord. IP1} & \textbf{Coord. IP2} & \textbf{Freq.} & \textbf{Cat.} & \textbf{Country} & \textbf{VPN} & \textbf{CEM flag} \\
\midrule
\cellcolor{red!20}(37.2980, 127.0777) & \cellcolor{green!20}(31.2222, 121.4581) & 34 & Porn & South Korea/China & No/No & No/No \\
\cellcolor{orange!20}(37.5245, 127.0354) & \cellcolor{red!20}(37.2980, 127.0777) & 30 & Porn & South Korea/South Korea & No/No & No/No \\
\cellcolor{red!20}(37.2980, 127.0777) & \cellcolor{blue!20}(24.1440, 120.6844) & 25 & Porn & South Korea/Taiwan & No/No & No/No \\
\cellcolor{blue!20}(24.1440, 120.6844) & \cellcolor{yellow!20}(37.4483, 127.0681) & 24 & Porn & Taiwan/South Korea & No/No & No/No \\
\cellcolor{blue!20}(24.1440, 120.6844) & \cellcolor{green!20}(31.2222, 121.4581) & 23 & Porn & Taiwan/China & No/No & No/No \\
\cellcolor{red!20}(37.2980, 127.0777) & \cellcolor{yellow!20}(37.4483, 127.0681) & 23 & Porn & South Korea/South Korea & No/No & No/No \\
\cellcolor{gray!20}(35.6893, 139.6899) & \cellcolor{red!20}(37.2980, 127.0777) & 23 & Porn & Japan/South Korea & Yes/No & No/No \\
\cellcolor{red!20}(37.2980, 127.0777) & (1.3521, 103.8200) & 22 & Porn & South Korea/Singapore & No/Yes & No/No \\
(48.4471, -123.3018) & \cellcolor{blue!20}(24.1440, 120.6844) & 22 & Porn & Canada/Taiwan & No/No & No/No \\
\cellcolor{orange!20}(37.5245, 127.0354) & \cellcolor{green!20}(31.2222, 121.4581) & 22 & Porn & South Korea/China & No/No & No/No \\
\bottomrule
\end{tabular}
}
\label{tab:top_ip_pairs}
\end{table}


Figure~\ref{fig:networktop} shows global high-frequency peer interactions and demonstrates our method's ability to isolate and analyze targeted subnets of interest. By using public metadata and flexible filtering criteria, we can create specialized network views that reflect specific behavioral patterns or risk indicators.


One example involves detecting IPs with known CEM distribution history. We must clarify that \textbf{this research does not include or process any torrents containing known CEM content}. Instead, some IPs that appeared in mainstream torrent swarms carry independent flags for prior CEM distribution activity. Their presence does not imply that they engaged in illegal activity within the samples we collected, but their behavioral overlap across different swarms offers an important risk signal. This finding demonstrates that torrent metadata can aid in identifying high-risk user patterns, even outside explicit CEM contexts.

\begin{figure}[h!]
    \centering
    \includegraphics[scale=0.35]{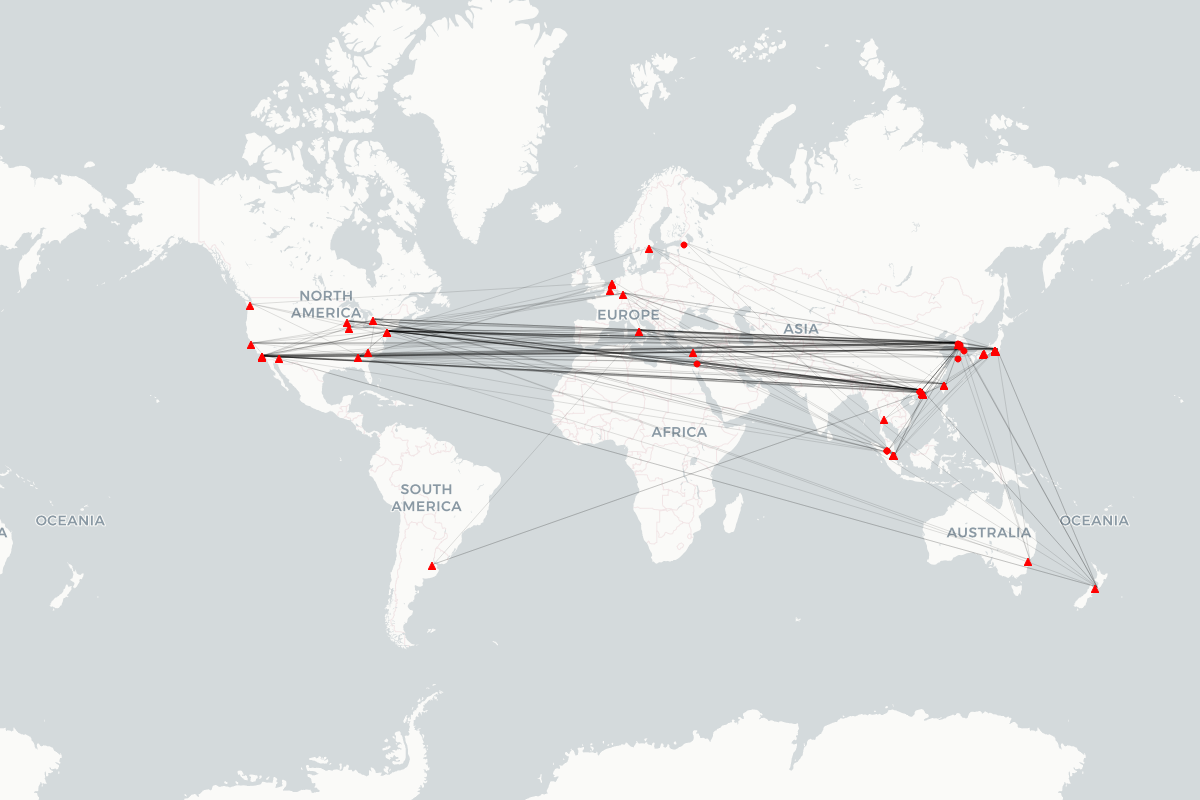}
    \caption{Network map of CEM-flagged IPs downloading exclusively pornographic content in our dataset.}
    \label{fig:network_cem_central}
\end{figure}

Figure \ref{fig:network_cem_central} shows a focused subnetwork of 70 CEM-flagged IP addresses that participated exclusively in pornographic torrents. Edges represent co-participation in the same swarm. This subnetwork is noticeably sparser than the broader network, with only 272 edges, but shows clear structure around several geographic clusters. High-density activity appears in East Asia, parts of Europe, and the United States. This pattern suggests that CEM-flagged users cluster within specific content niches, supporting the hypothesis that content type can help narrow behavioral profiling scope in OSINT applications.

We identified several highly influential nodes through betweenness centrality analysis that act as structural bridges in the network. These nodes connect otherwise weakly linked IP clusters, suggesting they play an important role in facilitating behavioral overlap or recurring swarm activity among users accessing pornographic content exclusively.

While the network spans globally, the most central nodes concentrate in North America, East Asia, and parts of Europe. This pattern shows regionally embedded activity hubs with strong international connections. Identifying both major hubs and smaller actors provides a useful basis for prioritizing investigation efforts and monitoring escalation risks.


Table~\ref{tab:cem_centrality} 
shows the ten most central IP addresses within the CEM-flagged pornographic network, ranked by betweenness centrality scores. We chose this metric because it identifies nodes that act as bridges between otherwise disconnected network parts, potentially revealing actors who could facilitate wider high-risk content spread. Each row includes approximate coordinates, country, centrality scores, VPN usage, and download counts. We added the CEM column as a reminder of the flag status.

\begin{table}[h!]
    \centering
    \caption[Top 10 IPs (coordinates) by betweenness centrality (CEM-flagged)]{Top 10 IPs (coordinates) by betweenness centrality (CEM-Flagged)}
    \label{tab:cem_centrality}
    \resizebox{\textwidth}{!}{
    \begin{tabular}{llrrrrrlrr}
\toprule
\textbf{Coordinates} & \textbf{Country} & \textbf{Betweenness} & \textbf{Degree} & \textbf{VPN} & \textbf{Downloads} & \textbf{CEM} \\
\midrule
(40.71230, -74.0068) & United States & 0.403 & 45 & True & 23 & Yes \\
(40.94680, 14.3015) & Italy & 0.170 & 24 & False & 13 & Yes \\
(43.89780, -79.2617) & Canada & 0.101 & 14 & False & 5 & Yes \\
(33.67500, -118.0027) & United States & 0.095 & 28 & False & 11 & Yes \\
(35.68930, 139.6899) & Japan & 0.087 & 6 & True & 3 & Yes \\
(37.53320, 126.9692) & South Korea & 0.086 & 9 & False & 4 & Yes \\
(37.52450, 127.0354) & South Korea & 0.080 & 26 & False & 10 & Yes \\
(33.74850, -84.3871) & United States & 0.066 & 20 & True & 9 & Yes \\
(1.28967, 103.8500) & Singapore & 0.053 & 10 & True & 3 & Yes \\
(37.53600, 126.9710) & South Korea & 0.052 & 11 & False & 5 & Yes \\
\bottomrule
\end{tabular}
    }
\end{table}


These metrics reveal both position and behavioral profile for each node. Unlike earlier analyzes, East Asia and the United States show the most dominant presence here, with South Korea and Japan contributing multiple high-centrality IPs. The highest-ranking IP combines high swarm participation with anonymization, potentially showing systematic behavior. In contrast, several other central IPs do not use anonymization, making them potentially more valuable for investigation since their associated metadata may be more reliable.

\begin{figure}[h!]
    \centering
    \includegraphics[width=\textwidth]{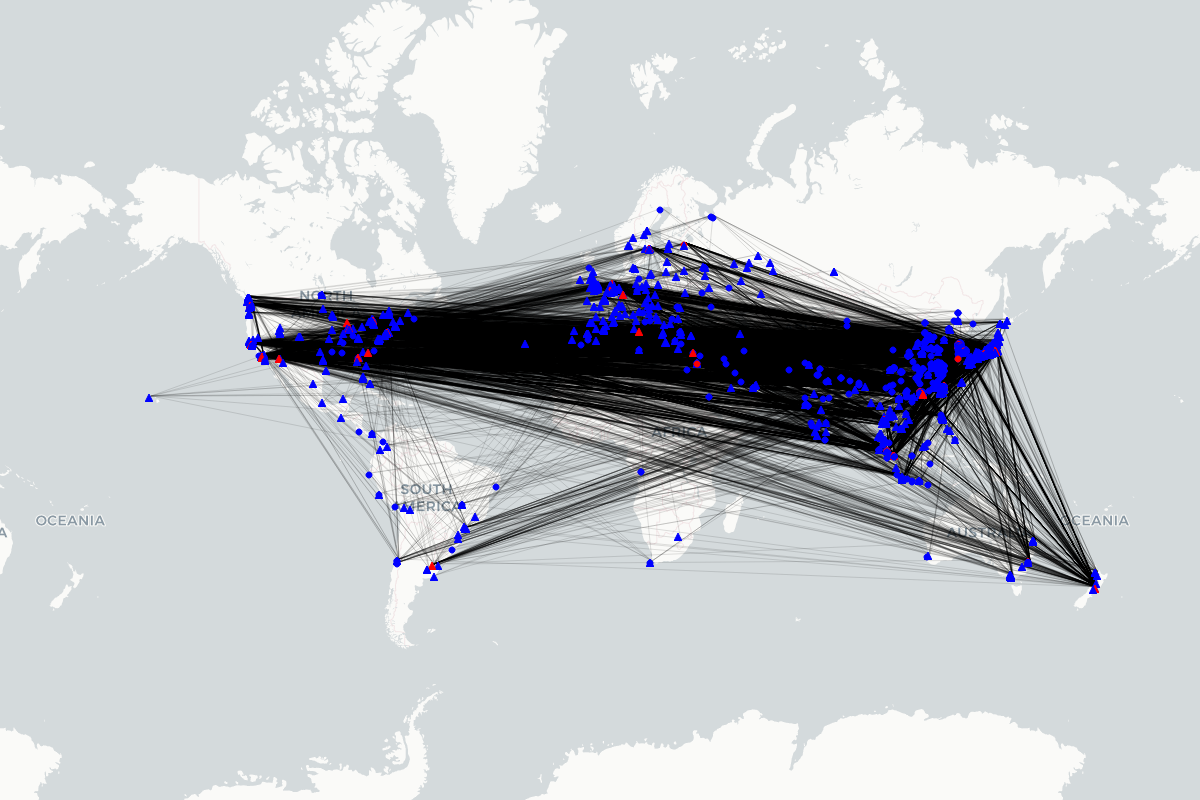}
    \caption[Extended network map of IPs distributing CEM]{Extended network map of IPs distributing CEM (only edges with flagged IPs shown)}
    \label{fig:network_cem_extended}
\end{figure}

Figure \ref{fig:network_cem_extended} expands the flagged IPs network by including their direct connections to non-flagged peers. This extended view shows only edges connected to flagged IPs; we excluded edges between non-flagged IPs for clarity. The full network contains 3,354 IPs and 374,327 connections. While Figure \ref{fig:network_cem_central} focused solely on flagged nodes, this updated view incorporates their immediate neighbors, who also participated exclusively in popular pornographic torrents. This selection enables us to examine the behavioral overlap between flagged users and participants of mainstream adult content.

To support this visualization, Table \ref{tab:cem_centrality_extended} 
presents updated centrality metrics for the extended network. While Figure \ref{fig:network_cem_extended} displays a filtered subset of edges (only those involving flagged IPs), Table \ref{tab:cem_centrality_extended}  
incorporates calculations based on the complete network, including edges between non-flagged IPs.

\begin{table}[h!]
\centering
\caption[Top 10 Coordinates by betweenness centrality (CEM-Flagged, extended network)]{Top 10 Coordinates by betweenness centrality (CEM-Flagged, extended network)}
\label{tab:cem_centrality_extended}
\captionsetup{format=plain}
\caption*{\small\textit{Note:} Cell colors indicate recurring IP addresses based on coordinates. Identical colors reflect the same IP across this table and Table \ref{tab:top_ip_pairs}}
\scalebox{0.80}{
\begin{tabular}{l l r r c r c}
\toprule
\textbf{Coordinates} & \textbf{Country} & \textbf{Betweenness} & \textbf{Degree} & \textbf{VPN} & \textbf{Downloads} & \textbf{CEM} \\
\midrule
\cellcolor{red!20}(37.2980, 127.0777) & South Korea & 0.067 & 2819 & False & 51 & No \\
\cellcolor{green!20}(31.2222, 121.4581) & China       & 0.047 & 2499 & False & 41 & No \\
\cellcolor{orange!20}(37.5245, 127.0354) & South Korea & 0.028 & 2171 & False & 34 & No \\
\cellcolor{gray!20}(35.6893, 139.6899) & Japan       & 0.027 & 1930 & True  & 32 & No \\
\cellcolor{yellow!20}(37.4483, 127.0681) & South Korea & 0.022 & 2127 & False & 29 & No \\
(47.6034, -122.3414) & United States & 0.021 & 1710 & True  & 17 & No \\
\cellcolor{blue!20}(24.1440, 120.6844) & Taiwan      & 0.018 & 2105 & False & 32 & No \\
(31.2222, 121.4581) & China       & 0.016 & 1744 & False & 22 & No \\
(1.3321, 103.8940)  & Singapore   & 0.015 & 1493 & True  & 18 & No \\
(37.4897, 127.0639) & South Korea & 0.013 & 1756 & True  & 24 & No \\
\bottomrule
\end{tabular}

}
\end{table}


Compared to Table~\ref{tab:cem_centrality}, 
which shows only CEM-flagged IP interactions, the extended table reveals substantial shifts in network influence. All top-ranking nodes by betweenness centrality are now non-flagged users, almost exclusively located in East Asia, who act as dense local hubs within pornographic swarms. However, the maximum betweenness centrality drops significantly in this denser network.


Many high-centrality IPs also appear in Table \ref{tab:top_ip_pairs},   
which captures IP pairs with the most co-downloads. Their recurring presence (marked by consistent cell coloring) shows they are both structurally central and behaviorally prominent. They sit at connection crossroads and engage repeatedly in shared downloads with other users, suggesting they act as core participants or behavioral anchors within their respective swarms.


Figure \ref{fig:network_cem_extended} shows how CEM-flagged users remain embedded in swarms with non-flagged users. Both groups share participation in exclusively pornographic torrents, indicating aligned content preferences. This pattern raises important questions about exposure pathways and user clustering mechanisms within torrent ecosystems. The global connection distribution, now extending more visibly into South America and West Asia, shows how sensitive content can spread across disconnected user bases through shared swarm activity.


While CEM distribution provides a clear marker of high-risk activity, it is not the only indicator of suspicious behavior. Our dataset also includes torrents flagged for sensitive or harmful material, notably books uploaded by `\textit{crwildman}', representing another important vector for risk assessment within intelligence frameworks.


To explore patterns beyond user interactions, we shift focus to the content side of the bipartite graph. We build a content-content network where nodes represent individual torrents and edges indicate shared downloaders, offering complementary insights into file relationships based on user behavior.

\begin{figure}
    \centering
    \includegraphics[width=\textwidth]{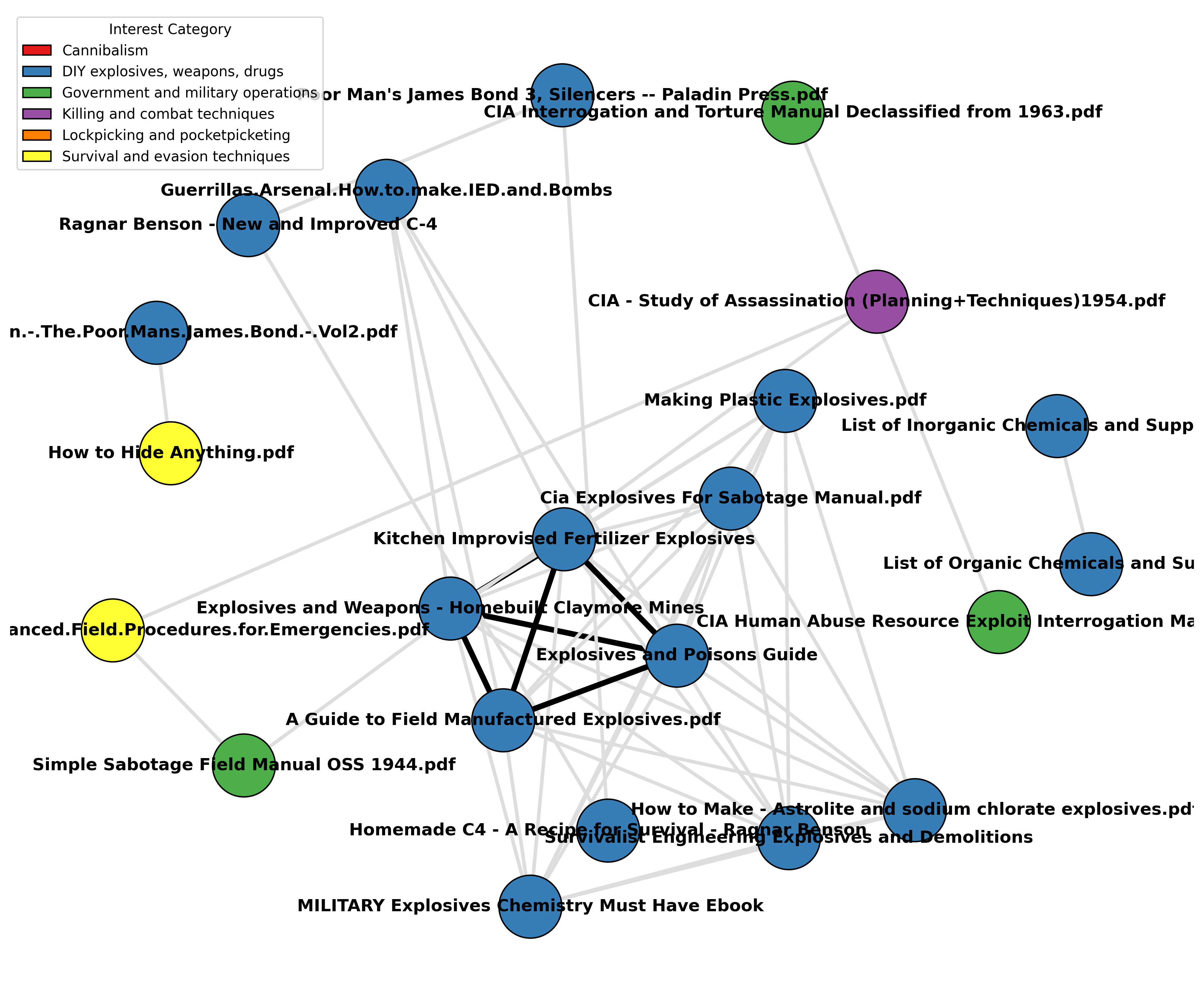}
    \caption{Content-to-content network for books uploaded by the user '\textit{crwildman}'}
    \label{fig:crwildman_content_network}
\end{figure}

Figure \ref{fig:crwildman_content_network} shows the resulting content-content network. Each node represents a torrent with potentially high-risk material, with edges denoting co-download relationships. We color-code nodes by thematic category, while edge color and thickness represent co-download frequencies.

The network shows dense clusters around explosives-related content, suggesting strong thematic connections among those users. Smaller groups focus on survival guides, assassination manuals, and chemical supply lists, which points to cross-topic interests for some users. Degree and betweenness centrality highlight four central nodes within the network. Table \ref{tab:crwildman_centrality} summarizes the most central titles, listing the number of unique peers, degree, and betweenness scores for each. You can match the connected titles to Table \ref{tab:connected_titles} for additional context.

\begin{table}[h!]
\caption{Sensitive book network metrics and peer discovery}
\centering
\scalebox{0.75}{
\begin{tabular}{p{0.5\textwidth}p{0.15\textwidth}p{0.09\textwidth}p{0.18\textwidth}p{0.25\textwidth}}
\toprule
\textbf{Title} & \textbf{Found peers} & \textbf{Degree} & \textbf{Betweenness} & \textbf{Connected titles (\#)} \\
\midrule
A Guide to Field Manufactured Explosives.pdf & 4 & 0.429 & 0.006 & 2, 3, 4, 5, 6, 7, 8, 9, 10 \\
Explosives and Poisons Guide & 2 & 0.429 & 0.006 & 1, 2, 4, 5, 6, 7, 8, 9, 10 \\
Explosives and Weapons - Homebuilt Claymore Mines & 4 & 0.429 & 0.006 & 1, 2, 3, 5, 6, 7, 8, 9, 10 \\
Kitchen Improvised Fertilizer Explosives & 2 & 0.429 & 0.006 & 1, 2, 3, 4, 5, 6, 8, 9, 10 \\
Cia Explosives For Sabotage Manual.pdf & 3 & 0.381 & 0.000 & 1, 3, 4, 6, 7, 8, 9, 10 \\
\bottomrule
\end{tabular}}
\label{tab:crwildman_centrality}
\end{table}

\vspace{0.5cm}

\begin{table}[h!]
\caption{Reference list of connected titles by number}
\centering
\scalebox{0.8}{
\begin{tabular}{p{0.03\textwidth}p{0.8\textwidth}}
\toprule
\textbf{\#} & \textbf{Connected Title} \\
\midrule
1 & A Guide to Field Manufactured Explosives.pdf \\ 
2 & Cia Explosives for Sabotage Manual.pdf \\ 
3 & Explosives and Poisons Guide \\
4 & Explosives and Weapons - Homebuilt Claymore Mines \\
5 & Guerrillas.Arsenal.How.to.make.IED.and.Bombs \\
6 & How to Make - Astrolite and sodium chlorate explosives.pdf \\
7 & Kitchen Improvised Fertilizer Explosives \\
8 & Making Plastic Explosives.pdf \\ 
9 & MILITARY Explosives Chemistry Must Have Ebook \\
10 & Survivalist Engineering Explosives and Demolitions \\
11 & Aquaman.and.the.Lost.Kingdom.2023.1080p.WEBRip.1400MB.DD5.1.x264 \\ 
\bottomrule
\end{tabular}}
\label{tab:connected_titles}
\end{table}


These titles show high degree centrality scores, meaning users who download them also tend to download multiple other sensitive documents rather than accessing them alone. These documents appear frequently in co-download patterns as part of broader interest clusters rather than as isolated downloads.


Additionally, the non-zero betweenness scores for most files suggest they serve as bridges between different network areas. Certain documents may act as entry points or crossroads for users exploring wider arrays of high-risk content. By contrast, ``\textit{Cia Explosives For Sabotage Manual.pdf}'' has lower degree and zero betweenness scores, showing that a more limited user group accesses it without connecting different clusters.

\begin{figure}
    \centering
    \includegraphics[scale=0.5]{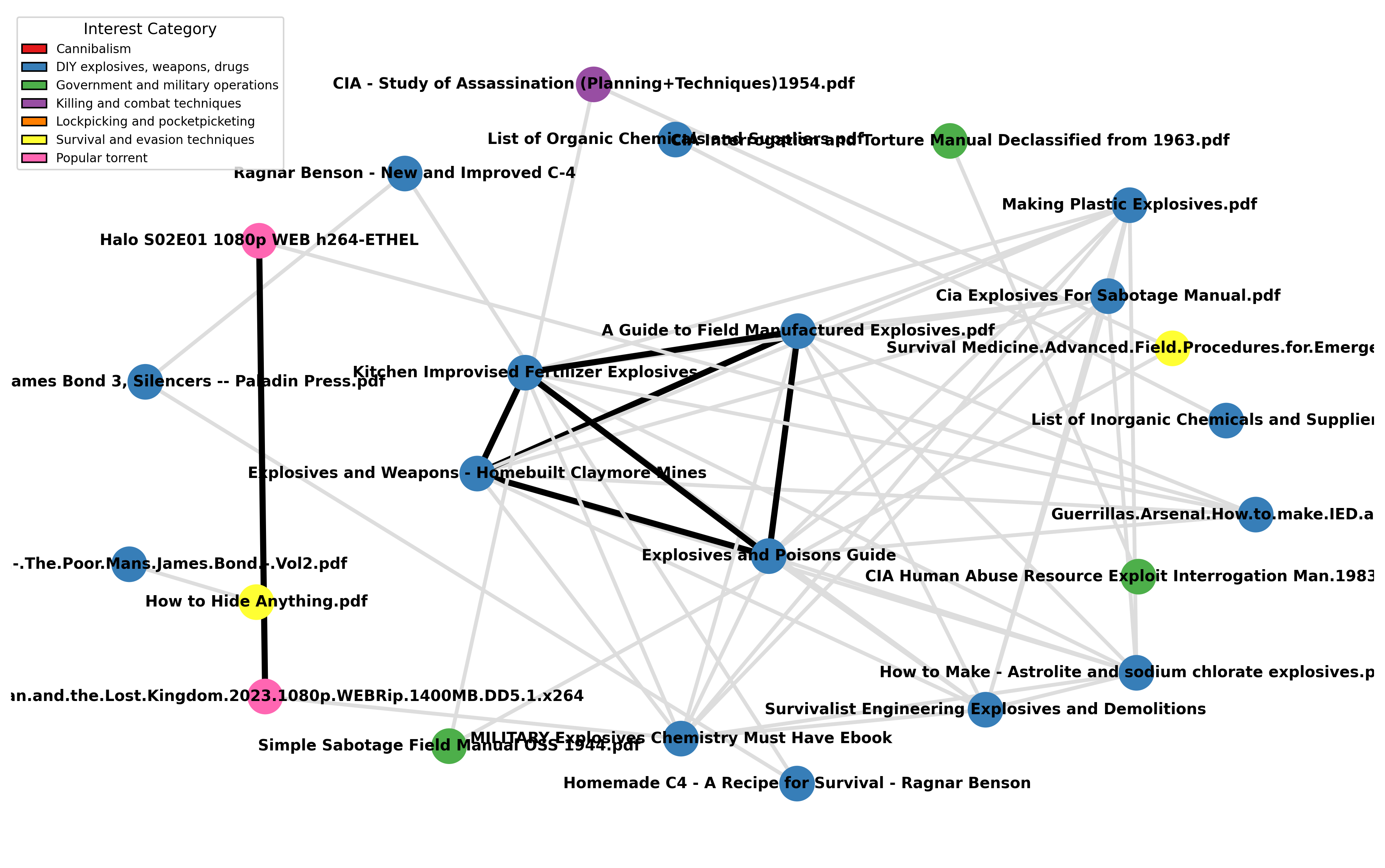}
    \caption{Extended content-to-content network}
    \label{fig:crwildman_extended_network}
\end{figure}

To better understand user behavior, we expanded the content-content network by adding popular torrents downloaded by the same users. This broader network (Figure \ref{fig:crwildman_extended_network}) links high-risk documents to mainstream media downloads, providing deeper insights into user interests.

Figure \ref{fig:crwildman_extended_network} reveals several layers of actionable insights. First, it helps identify users whose download patterns link both high-risk documents and mainstream entertainment, creating behavioral fingerprints. The connection between sensitive files and popular torrents allows for profiling based on thematic co-consumption.


Furthermore, the network also highlights engagement differences: some users download multiple sensitive files and a range of popular torrents, while others are single-point connections, likely casual or one-off exposures. Clusters in the network show thematic overlaps, such as between guerrilla literature and military-themed entertainment (e.g.,  between \textit{Halo S02E01 1080p WEB h264-ETHEL} and \textit{Guerrillas.Arsenal.How.to.make.IED.and.Bombs}). These patterns enable profiling based on thematic co-consumption behavior, providing a richer, multi-dimensional view of risk than sensitive content downloads alone.

To complement the visual insights of Figure \ref{fig:crwildman_extended_network}, Table \ref{tab:crwildman_extended_centrality} presents network metrics for the key torrents within the extended content-to-content network, using the same structure as Table \ref{tab:crwildman_centrality}.

\begin{table}[h!]
\caption{Extended network: document metrics and peer discovery.}
\centering
\scalebox{0.75}{
\begin{tabular}{p{0.5\textwidth}p{0.1\textwidth}p{0.09\textwidth}p{0.18\textwidth}p{0.25\textwidth}}
\toprule
\textbf{Title} & \textbf{Found peers} & \textbf{Degree} & \textbf{Betweenness} & \textbf{Connected titles (\#)} \\
\midrule
MILITARY Explosives Chemistry Must Have Ebook & 4 & 0.391 & 0.035 & 1, 2, 3, 4, 6, 7, 8, 10, 11\\
Kitchen Improvised Fertilizer Explosives & 2 & 0.391 & 0.008 & 1, 2, 3, 4, 5, 6, 8, 9, 10 \\
A Guide to Field Manufactured Explosives.pdf & 4 & 0.391 & 0.008 & 2, 3, 4, 5, 6, 7, 8, 9, 10 \\
Explosives and Poisons Guide & 2 & 0.391 & 0.008 & 1, 2, 4, 5, 6, 7, 8, 9, 10\\
Explosives and Weapons - Homebuilt Claymore Mines & 4 & 0.391 & 0.008 & 1, 2, 3, 5, 6, 7, 8, 9, 10\\
\bottomrule
\end{tabular}}
\label{tab:crwildman_extended_centrality}
\end{table}


The metrics confirm earlier findings: documents about improvised explosives rank highest in degree and betweenness centrality, repeatedly forming the core of both focused and extended networks. Their central roles highlight their appeal and thematic importance. Notably, \textit{MILITARY Explosives Chemistry Must Have Ebook} stands out as the only top-ranked document directly linked to a popular torrent, suggesting some users engage with both high-risk instructional content and mainstream media like DC Comics. Such overlaps help refine user profiles by capturing multi-interest download patterns.
Combining torrent metadata with network analysis improves risk assessment, escalation detection, and investigative prioritization, supporting privacy-preserving OSINT methods.



While broader network analysis reveals general escalation risks, outliers provide compelling investigative signals. As Figure \ref{fig:crwildman_geo_distribution} shows, one IP address in Mumbai stands out for its unusually high engagement—this user downloaded nine books focused on improvised explosives yet did not access popular torrents suggesting a deliberate interest in high-risk material rather than casual file sharing. Their activity, combined with residential ISP affiliation (\textit{Space Vision Digital Network Pvt. Ltd.}) and geographic specificity, indicates heightened escalation risk that warrants further scrutiny.
These results show how openly available torrent metadata enables behavior-based segmentation and strengthens P2P profiling. Contextual metadata and nuanced filtering are key for risk assessment, early warning, and prioritizing investigations using open-source data.


The next section shifts focus to torrent metadata's technical value for OSINT workflows. We evaluate its suitability and efficiency, covering availability, structure, reliability, and processing workload.


Both torrent and IP metadata proved highly available, with \textit{The Pirate Bay} online and trackers providing many IPs. IP metadata sources allow reliable geolocation and ISP lookup, supporting real-time OSINT.


Torrent metadata is semi-structured—fields like category, uploader, and magnet link are consistent, but file titles vary by uploader. This limits automation, so preprocessing is needed. Hashes alone add little context, and tracker queries are often unreliable for missing data.


IP metadata varies more in structure: city, country, ISP, and AS fields are usually present and geographic coordinates help mitigate textual inconsistencies. Despite occasional errors, IP metadata remains reliable for geospatial intelligence. However, anonymization technologies reduce geolocation accuracy; about 20\% of the sampled IPs use privacy-preserving services. Some VPN use may be underreported, so we checked ISP/AS fields for privacy signals. Still, reputable providers deliver mostly reliable data, though anonymized activity remains a key limitation.


Scraping torrent metadata was efficient with tools like Selenium and BeautifulSoup. UDP tracker queries require binary socket programming and robust parsing; however, but our system is reusable. Processing many torrents is time-consuming, and peer lists can be incomplete; thus, so regular scraping ensures higher accuracy.



IP metadata extraction required extra post-processing, as public APIs (like \textit{ip-api.com} and \textit{ip-info.io}) impose rate limits and variability, so fallback or paid options are needed for scaling.

In summary, torrent and IP metadata offer a technically feasible but complex data source for OSINT. Their reliability and structure vary, making flexible extraction systems essential for robust analysis.
\section{Conclusions}\label{chapter:conclusions}


This study examined how torrent metadata can systematically support OSINT workflows aimed at identifying illegal content and users distributing high-risk material. Through a proof-of-concept pipeline, we demonstrated how publicly available torrent metadata—such as tracker responses, IP geolocation, and cross-source flagging—can reveal behavioral patterns in decentralized and partly anonymized networks. By integrating these metadata sources, the approach enhances user profiling, risk identification, and investigative prioritization while maintaining compliance with privacy-preserving analytical standards.

The central research question that guided this work was as follows:
\begin{quote}
\textbf{RQ:} To what extent does our proof of concept demonstrate the potential of torrent metadata for OSINT-based user profiling?
\end{quote}

\begin{framed}
The proof of concept demonstrates that torrent metadata can be effectively leveraged for OSINT-based user profiling. It enables the detection of behavioral patterns, user clusters, and potential high-risk signals based on swarm participation, anonymization use, and content interaction. 
\end{framed}



The pipeline mapped over 60,000 unique IP addresses to specific torrents by querying public trackers and collecting peer data, enabling user–content correlation, behavioral network construction, and detection of repeated or anomalous activity. Graph-based analysis identified co-download patterns and thematic clusters, linking content types to user groups. Despite limitations such as anonymization and partial swarm visibility, the findings confirm that torrent metadata holds strong potential as an OSINT source when processed and contextualized. 
The research also addressed three sub-questions: defining torrent metadata and its retrievability (sRQ1), examining structural relationships between users and content through network mapping (sRQ2), and evaluating the data’s suitability for real-world OSINT applications (sRQ3). Below each question is systematically explored to demonstrate the feasibility and operational relevance of this approach.

\begin{quote}
    \textbf{sRQ1:} What type of data is available, retrievable, and usable for detecting high-risk behavior?
\end{quote}

\begin{framed}
Publicly accessible torrent-related data, including content metadata, swarm IPs, and IP-level attributes, can be retrieved and structured for behavioral profiling, particularly when enriched with cross-source signals.
\end{framed}



We collected torrent metadata from public index sites, trackers, and IP intelligence services, including torrent titles, tags, file structures, swarm peer IPs, geolocation, organizational data, and anonymization indicators. For flagged content such as CEM and weapons manuals, repeated participation patterns emerged, showing that some user behaviors remain trackable through swarm activity. These findings confirm the richness and retrievability of torrent data. Querying public UDP trackers enabled real-time mapping of torrents to active peer IPs, allowing content–user correlation. However, uncovering deeper behavioral relationships requires moving beyond isolated activity toward network-level analysis, leading to sRQ2.

\begin{quote}
   \textbf{sRQ2:} How effectively can our proof of concept map torrent user networks using publicly available data? 
\end{quote}

\begin{framed}
The proof of concept effectively mapped user networks, revealing structural relationships, behavioral clusters, and central actors within swarms based on co-download patterns and shared content interest.
\end{framed}


Using unipartite graph projections and co-download logic, we built user and content networks capturing behavioral links. IPs frequently co-occurring in swarms tended to cluster together, revealing potential collaboration, shared interests, or routines. Some nodes connected both high-risk (e.g., weapons manuals) and benign (e.g., military-themed TV) downloads, indicating thematic overlap warranting further profiling. Centrality analysis identified recurring IPs with broad swarm participation as potential hubs or repeat offenders. This approach showed how behavioral patterns emerge from torrent metadata when analyzed through network projections.



These findings show that network-based approaches are viable for profiling and escalation detection in OSINT contexts. We showed and discussed how torrent data can reveal hidden structures among users and content. The data can be used not just for individual profiling but also for uncovering group-level dynamics. To fully assess the feasibility of this approach for real-world OSINT, we must now evaluate the overall suitability of the data, addressed in sRQ3.

\begin{quote}
    \textbf{sRQ3:} How suitable and efficient are torrent and IP metadata as sources for OSINT, in terms of availability, structure, reliability, and processing effort?
\end{quote}

\begin{framed}
Torrent and IP metadata are generally suitable and structurally consistent for OSINT use, though challenges such as anonymization, geolocation inaccuracies, and data completeness must be mitigated.
\end{framed}




Torrent and IP metadata are generally accessible and structurally suitable for OSINT analysis but face challenges such as data noise, inconsistent completeness, and limitations in IP geolocation. About 20\% of IPs use VPNs or proxies, complicating attribution; yet, anonymity detection itself serves as a valuable OSINT signal. Among IPs linked to CEM distribution, anonymization exceeds 75\%, marking a clear behavioral divide useful for prioritizing threats. Anonymization usage, especially when correlated with sensitive content, offers insight into user intent and sophistication, helping investigators focus on higher-risk profiles.
Despite technical limitations such as tracker reliance and API constraints, torrent and IP metadata are obtainable via automation, consistent enough for graph analysis, and provide reliable geolocation, organizational, and privacy signals. These strengths support integrating such data into scalable OSINT pipelines for investigative use.

\section{Limitations and Future works}

Despite promising results, several limitations constrain the current research:
\begin{enumerate}
    \item \textbf{Manual and non-scalable data collection}: The scraping and analysis pipeline required manual setup and intervention, limiting scalability for real-time or long-term monitoring.
    \item \textbf{Incomplete swarm visibility}: Relying solely on UDP tracker responses excludes peers participating only via DHT, which may lead to underreporting of swarm composition.
    \item \textbf{Anonymization obfuscation}: Around 20\% of observed IPs used privacy-preserving technologies, and some anonymised IPs likely went undetected due to limitations in metadata sources and detection logic.
    \item \textbf{Limited topical scope}: The research case study deliberately limited its scope to a single uploader and one keyword category. This methodological choice was made to minimize the risk of accidental exposure to illegal content for all members and components involved in the research project.    
    \item \textbf{Limited keyword}: Broader keyword sampling could offer more robust behavioral generalization.
    \item \textbf{Temporal limitations}: Torrent data are time-sensitive. Without continuous scraping, longitudinal behavior or escalation cannot be tracked accurately.
\end{enumerate}


In addition to these technical constraints, privacy implications form a critical limitation of this research. Although participation in torrent swarms exposes IP addresses, these are not directly linked to personally identifiable information. Most IPs, especially dynamic ones, identify only a network endpoint, not a user. Linking an IP to an individual would require access to ISP subscriber records, which is beyond the scope of this OSINT-based research.




However, it is important to note that this study only aims to provide proof of concept, demonstrating that behavioral profiling and partial de-anonymization are feasible in the context of criminal torrent sharing data. By integrating additional OSINT data, even users behind VPNs may be de-anonymized, though this remains a more complex process.
This work does not aim to definitively identify individuals but instead focuses on profiling behavioral patterns and detecting signals of high-risk activity observable at the network level.

\section{Future work}
As already mentioned and discussed, this work intend to present a proof of concept approach to profile and highlight illegal behaviors in the context of torrent data sharing. However, to enhance the capabilities and relevance of this approach, future research should consider

\begin{enumerate}
    \item \textbf{Automation and scaling}: While the current implementation relies on manual batch-based scraping and post-processing, future works could incorporate real-time data collection from torrent trackers, automated IP enrichment, and live network graph generation. This would allow the profiling process to scale and adapt dynamically to emerging threats.
    \item \textbf{DHT integration}: The reliance on UDP tracker scraping could be complemented with Distributed Hash Table (DHT) queries to increase peer visibility, especially for trackerless torrents. DHT-based discovery would provide a more comprehensive swarm coverage and reduce reliance on potentially throttled trackers.
    \item \textbf{Cross-source OSINT fusion}: Future implementations could integrate richer IP-level data (e.g., historical activity, network signatures, device fingerprints) or external flagging sources to improve the accuracy of risk assessment. Behavioral features such as download timing, frequency, and cross-torrent navigation could also strengthen user profiles. 
    \item \textbf{Prioritization and scoring mechanisms}: Introducing a scoring system to rank user risk based on swarm participation, anonymization behavior, and content sensitivity could help process large datasets and support investigative prioritization.
    \item \textbf{Extending case study design}: Building additional and larger case studies focused on other types of high-risk content would help validate the generalizability of the pipeline across domains and provide insight into evolving threat ecosystems.
    \item \textbf{Ethical framework development}: Given the sensitive nature of profiling, develop a privacy-aware framework to ensure ethical handling of open-source data in investigative contexts.
\end{enumerate}

Our proposed OSINT platform represents only a building block of a more complex platform. In the future, we plan to crawl information from the dark web to search for specific keywords related to illegal content. 
Specifically, we know that Pirate Bay is a free, searchable online index of different types of data (video, audio, books, etc.) and does not contain explicit material. However, Btdigg is another well-known torrent sharing platform. Users on the dark web use this uncensored platform to share CEM materials. However, to intercept this specific critical content, we must first crawl for keywords from forums, Telegram channels, and chats, and then use these findings in our proposed platform.

\section*{Abbreviations}\label{Abbreviations}

\addcontentsline{toc}{chapter}{Abbreviations}

\begin{tabular}{ll}
API         & Application Programming Interface \\
AS          & Autonomous System \\
BEA         & Behavioral Evidence Analysis \\
CEM         & Child Exploitation Material \\
CSV         & Comma-Separated Values \\
DHT         & Distributed Hash Table \\
DIY         & Do It Yourself \\
IP          & Internet Protocol \\
ISP         & Internet Service Provider \\
OSINT       & Open Source Intelligence \\
P2P         & Peer-to-Peer \\
PoC         & Proof of Concept \\
RQ          & Research Question \\
sRQ         & Sub-Research Question \\
TCP         & Transmission Control Protocol \\
UDP         & User Datagram Protocol \\
VPN         & Virtual Private Network \\
\end{tabular}

\bibliographystyle{elsarticle-num}
\bibliography{references.bib}



\end{document}